\documentstyle[12pt,cite,epsfig]{article} 

\begin{document}

~\hspace*{5.0cm} ADP-AT-00-1\\
~\hspace*{5.cm} {\it Astropart. Phys.}, revised\\

\begin{center}
{\large \bf A Proton Synchrotron 
Blazar Model for Flaring in Markarian~501}\\[1cm] 
A. M\"ucke\footnote{present
address: Universit\'e de Montr\'eal, D\'epartement de Physique,
Montr\'eal, H3C 3J7, Canada} and R.J. Protheroe\\Department of
Physics and Mathematical Physics\\ The University of Adelaide,
Adelaide, SA 5005, Australia\\[2cm]
\end{center}


\begin{abstract}
The spectral energy distribution (SED) of gamma-ray loud BL~Lac objects typically has a
double-humped appearance usually interpreted in terms of
synchrotron self-Compton models.  In proton blazar models, the
SED is instead explained in terms of acceleration of protons and
subsequent cascading.  We discuss a variation of the Synchrotron
Proton Blazar model, first proposed by M\"ucke \& Protheroe
(1999), in which the low energy part of the SED is mainly
synchrotron radiation by electrons co-accelerated with protons
which produce the high energy part of the SED mainly as
proton
synchrotron radiation.

As an approximation, we assume non-relativistic shock acceleration
which could apply if the bulk of the plasma in the jet frame were
non-relativistic. Our results may therefore change if a relativistic
equation of state were used. We
consider the case where the
maximum energy of the accelerated protons is above the threshold
for pion photoproduction interactions on the synchrotron photons
of the low energy part of the SED. Using a Monte Carlo/numerical
technique to simulate the interactions and subsequent cascading
of the accelerated protons, we are able to fit the high-energy
gamma-ray portion of the
observed SED
of Markarian 501 during the April 1997 flare.  We find that the
emerging cascade spectra initiated by gamma-rays from $\pi^0$
decay and by $e^\pm$ from $\mu^\pm$ decay turn out to be
relatively featureless.  Synchrotron radiation produced by
$\mu^\pm$ from $\pi^\pm$ decay, and even more importantly by
protons, and subsequent synchrotron-pair cascading, is able to
reproduce well the high energy part of the SED.  For this fit we
find that synchrotron radiation by protons dominates the TeV
emission, pion photoproduction being less important with the
consequence that we predict a lower neutrino flux than in other
proton blazar models.
\end{abstract}


\hspace*{2cm} Gamma-rays: theory, Neutrinos, Synchrotron emission, Cascade simulation

\date{Submitted to Astroparticle Physics, April 2000}

\section{Introduction}

During its giant outburst in April 1997, the nearby BL~Lac object
Mkn~501 (at redshift z=0.034) emitted photons up to $24$~TeV and
$0.5$~MeV in the $\gamma $-ray and X-ray bands, respectively, and
has proved to be the most extreme TeV-blazar observed so far
(e.g. Catanese et al 1997, Pian et al 1998, Protheroe et al 1998,
Quinn et al 1999, Aharonian et al 1999). This energy is the
highest so-far observed for any BL~Lac object, and the flux is
approximately 2 orders of magnitude higher than the synchrotron
peak at its quiescent level. BeppoSAX and OSSE observations
(Maraschi 1999) suggest that the X-ray spectrum is curved at all
epochs, and the spectrum during flaring has been fitted by a
multiply-broken power-law (Bednarek \& Protheroe 1999). COMPTEL
has not seen any significant signal from Mkn~501 at any time
(Collmar 1999), while a $3\sigma $ upper limit of $F(>100\,{\rm
MeV})<3.6\times 10^{-7}$cm$^{-2}$ s$^{-1}$ has been derived for
the April 1997 EGRET viewing period (Catanese et al 1997).

A flux increase at TeV-energies was also observed with the
Whipple, HEGRA and CAT telescopes (Catanese et al 1997), with the
most intense flare peaking on April 16 at a level $\sim100 $
times higher than during its quiescent flux. The non-detection of
Mkn~501 by EGRET indicates that most of the power output of the
high energy component is in the GeV-TeV range. The
TeV-observations revealed a power-law spectrum with photon index
$\sim 2$ up to $\sim 10$~TeV and a gradual steepening up to
24~TeV. The extragalactic diffuse infrared background leads to
significant extinction of $\gamma $-rays through $\gamma \gamma
$-pair production above 10~TeV. The extinction-corrected
TeV-spectrum (e.g. Bednarek \& Protheroe 1999), shows the
spectral energy distribution (SED) peaking at $\sim 2$~TeV.
Optical observations did not show any significant variations
(Buckley \& McEnery 1997), indicating that the change in the low
energy part of the SED was mainly confined to the X-ray band above
$0.1$ keV.

Various models have been proposed to explain the observed $\gamma
$-ray emission from TeV-blazars, all of which are identified as
high-frequency peaked BL~Lac objects. Leptonic models, in which
electrons inverse-Compton scatter a population of low energy
photons to high energies, currently dominate the thinking of the
scientific community. Because of the low luminosity of accretion
disks in BL~Lacs, the main target photons for the relativistic
electrons would be the synchrotron photons produced by the same
relativistic electron population, as in the synchrotron
self-Compton (SSC) model. An alternative scenario for the
production of the observed $\gamma $-ray flux has been proposed
involving pion photoproduction by energetic protons
with
subsequent synchrotron-pair cascades initiated by decay products
(photons and $e^\pm$) of the mesons
(e.g. Mannheim et al 1991, Mannheim 1993). These proton-initiated cascade (PIC) 
models could, in principle, be
distinguished by the observation of high energy neutrinos
produced as a result of photoproduction.

In this paper, we consider the April 1997 flare of Mkn~501 in the
light of a modified Synchrotron Proton Blazar (SPB) model. We
assume that electrons ($e^{-}$) and protons ($p$) are accelerated
by 1st order Fermi acceleration at the same shock. The
relativistic $e^{-}$ radiate synchrotron photons which serve as
the target radiation field for proton-photon interactions, and
for the subsequent pair-synchrotron cascade which develops as a
result of photon-photon pair production. This cascade
redistributes the photon power to lower energies where the
photons escape from the emission region, or ``blob,'' which moves
relativistically in a direction closely aligned with our
line-of-sight.

Until recently, this model was not able to reproduce the general
features of the double-humped blazar spectral energy distribution
(SED), but produced a rather featureless spectrum (see e.g.
Mannheim 1993), nor could it explain correlated X-ray/TeV-variability.
Here, we present a comprehensive description of
our Monte-Carlo simulations of a stationary SPB model, including
all relevant emission processes, and show that this model is
indeed capable of reproducing a double-humped SED as
observed. Here, the origin of the TeV-photons are proton
synchrotron radiation, as first proposed by M\"ucke \& Protheroe
(1999); a similar model has also been proposed by Aharonian
(2000), and Rachen (1999) presented speculations about
$\mu^\pm$- and proton-synchrotron radiation leading to narrow
cascade spectra during flares, which might explain correlated
X-ray/TeV-variability.
  Jet energetics and limits from particle shock
acceleration, however, put severe constraints on this
scenario. The goal of this paper is to discuss the physical
processes included in our SPB model Monte-Carlo code, and give
the results of applying this code, as an example, to reproduce
the SED of the giant flare from Mkn~501 which occurred in April
1997. A comprehensive study of the whole parameter-space
(magnetic field, Doppler factor, etc.) for this model will be the
subject of a subsequent paper.

In Section 2, we discuss constraints on the maximum particle
energies imposed by the co-acceleration scenario, and by the pion
production threshold. Section 3 is devoted to the emission
processes in the present model. Energy losses and particle
production are treated in Sect. 3.1, while the cascade
calculations, including a brief description of our code, are
outlined in Sect. 3.2. In Sect. 4 we apply our model to the April
1997 flare of Mkn~501. The multifrequency photon spectrum is
shown in Sect. 4.1, while in Sect. 4.2 the predicted neutrino
spectrum is discussed. We conclude with a discussion and summary
in Section 5.


\section{The Co-acceleration Scenario}
In the present model, shock accelerated protons ($p$) interact in
the synchrotron photon field generated by the electrons ($e^{-}$)
co-accelerated at the same shock. This scenario may put
constraints on the maximum achievable particle energies.

The usual process considered for accelerating charged particles in
high energy astrophysics is diffusive shock acceleration (see
e.g. Bell 1978, Drury 1983, Blandford \& Eichler 1987, Biermann \&
Strittmatter 1987, Jokipii 1987, Jones \& Ellison 1991), in which
particles undergo collisionless scattering, e.g.  by Alfv\'{e}n waves,
in the upstream and downstream plasma. Charged particles with
gyroradii larger than the thickness of the shock front propagate with
diffusion coefficients $\kappa _{1}$ and $\kappa _{2}$ in the upstream
and downstream plasma, respectively, for propagation parallel to the
shock normal. In the shock frame the plasma flow velocity changes from
$u_{1}=\beta_1 c$ in the upstream region to $u_{2}$ in the downstream
region.  In this paper, for simplicity we restrict ourselves to
non-relativistic shocks, and postpone discussion of relativistic shock
acceleration to a later paper.

The acceleration time scale for non-relativistic shocks is given
by
\begin{equation}
t_{\rm
acc}=\frac{3r_{c}\beta}{(r_{c}-1)u_{1}^{2}}(\kappa_{1}+r_{c}\kappa_{2})
\end{equation}
 where $r_{c}\rightarrow 4$ is the compression ratio for the case
of strong shocks in a non-relativistic monoatomic ideal gas. If
the magnetic field is governed by an ordered component, the
orientation of the shock normal to the main magnetic field
direction becomes important. In general, the diffusion
coefficient can be written as
\begin{equation}
\kappa_{i}=\kappa_{i,||}\cos^{2}\theta_{i}+\kappa_{i,\perp}\sin
^{2}\theta_{i},\hspace*{.5cm}i=1,2
\end{equation}
where $\theta_{i}$ is the angle between the magnetic field and
the axis connecting the upstream ($i=1$) and downstream ($i=2$)
regions. In the diffusion limit, kinetic theory relates the
parallel and perpendicular diffusion coefficients through
\begin{equation}
\kappa_{||}=[1+\eta^{2}]\kappa_{\perp}=\frac{1}{3}\lambda_{||}\beta c
\end{equation}
where $\eta =\lambda_{||}/{r_{g}}$ with $\lambda_{||}$ being the
mean free path parallel to the magnetic field, $r_{g}=
\beta\gamma m c^{2}/eB$ is the particle's gyroradius, $m$ and
$\gamma=(1-\beta^2)^{-{1/2}}$ are the particle's mass and Lorentz
factor, respectively, and $B$ is the magnetic field strength in
the upstream region.  The mean free path is, in general, a
function of the particle energy through its gyroradius, and is
dependent on the spectrum of the magnetic turbulence. In the
small angle scattering approximation (i.e., if Alfv\'{e}n waves
dominate the particle deflection with wavelength equal to the
particle gyroradius; see Drury 1983) we have
\begin{equation}
\lambda =\frac{B^{2}r_{g}}{8\pi I(k)k}.
\end{equation}
This spectrum $I(k)$ is usually expressed as a power law of the
wave number $k$ in the turbulent magnetic field:
\begin{equation}
I(k)\propto k^{-\delta }.
\end{equation}
$\delta =5/3$ corresponds to Kolmogorov turbulence which may be
common in astrophysical environments (Biermann \& Strittmatter
1987), while $\delta =1$ corresponds to Bohm diffusion, and is
often considered for simplicity. For strong magnetic fields,
Kraichnan turbulence $\delta =3/2$ (Kraichnan 1965) may be
present. In the following, we consider $\delta $ as a free
parameter.  The mean free path may then be expressed as (see
Biermann \& Strittmatter 1987)
\begin{equation}
\lambda _{||}=\frac{r_{g}}{b(\delta -1)}\left(
\frac{r_{g,{\rm max}}}{r_{g}}\right) ^{\delta -1} \hspace*{.3cm}\mbox{for}\hspace*{.3cm}\delta \neq 1
\end{equation}
\[
\lambda _{||}=\frac{r_{g}}{b}\left[\ln\left(
\frac{r_{g,{\rm max}}}{r_{g,{\rm min}}}\right)\right] ^{-1} \hspace*{.3cm}\mbox{for}\hspace*{.3cm}\delta=1
\]
where $b$ is the ratio of the turbulent to ambient magnetic energy
density, $r_{g,{\rm max}}$ is the gyroradius of the most energetic
protons, and has the same order of magnitude as the system size,
and $r_{g,{\rm min}}$ corresponds to the smallest turbulence scale.  The
mean free path, and consequently the acceleration time at maximum
energy is only slightly dependent on the turbulence spectrum in the
case of protons, whereas for electrons the acceleration time at
maximum energy shows a strong dependence on the magnetic turbulence
spectrum adopted.  

We expect $b\leq 1$, since otherwise the
energy density in particles would not be able to be confined by
the ambient field (Biermann \& Strittmatter 1987). 
With these relations, the acceleration time scale
may be re-written as
\begin{equation}
t_{\rm acc} = \frac{r_g \beta c}{u_1^2}
F(\theta_1,\eta)
\end{equation}
where
\begin{equation}
F(\theta_1,\eta) = \frac{\eta r_c}{r_c-1} \left[ \cos^2{\theta_1}
+ \frac{\sin^2{\theta_1}}{(1+\eta^2)} +
\frac{r_c\cos^2{\theta_1}+r_c^3\sin^2{\theta_1}/(1+\eta^2)}{(\cos^2{\theta_1}
+ r_c^2 \sin^2{\theta_1})^{3/2}} \right]
\end{equation}  
The diffusion approximation used here limits the maximum mean
free path to $\eta < \beta/\beta_1$ (Jokipii 1987).  If the
particle spectra are cut off due to synchrotron losses, balancing
the acceleration time scale with the loss time scale determines
the maximum Lorentz factors of protons,
\begin{equation}
\gamma_{p,\rm{max}} = 2.1 \times 10^{11}
\beta_1 \left[ \beta\, B\, F(\theta_1,\eta_{p,\rm{max}}) \right]^{-1/2}
\end{equation}
and electrons
\begin{equation}
\gamma_{e,\rm{max}} = 1.2 \times 10^{8}
\beta_1 \left[ \beta\, B\, F(\theta_1,\eta_{e,\rm{max}}) \right]^{-1/2}
\end{equation}
where $B$ is in Gauss.  The ratio of the
maximum proton Lorentz factor to the maximum electron Lorentz
factor is then
\begin{equation}
\frac{\gamma_{p,\rm{max}}}{\gamma_{e,\rm{max}}} \leq
\frac{m_p}{m_e} \left[{\frac{
F(\theta_1,\eta_{e,\rm{max}})}{
F(\theta_1,\eta_{p,\rm{max}})}}\right]^{1/2}
\label{eq:gammamax_ratio}
\end{equation}
where the equality corresponds to the maximum proton energy being
determined by synchrotron losses, and the inequality to the
maximum proton energy being determined instead by adiabatic
losses.  For parallel shocks this relation is consistent with the
results found by Biermann \& Strittmatter (1987).

The corresponding acceleration time scales at the maximum
particle energies, if determined by synchrotron losses, are
\begin{equation}
t_{\rm{acc},p,\rm{max}} = 2.2\times 10^7 B^{-3/2} \beta_1^{-1}
\left[\beta F(\theta_1,\eta_{p,\rm{max}}\right]^{1/2}
\;\;\;\; {\rm s}
\label{eq:tacc_p}
\end{equation}
for protons, and
\begin{equation}
t_{\rm{acc},e,\rm{max}} = 6.6 B^{-3/2}\beta_1^{-1}
\left[\beta F(\theta_1,\eta_{e,\rm{max}}\right]^{1/2}
\;\;\;\; {\rm s}
\end{equation}
for electrons.  In the present paper, we shall adopt
$\beta_1=0.5$, $\beta=1$ and $r_c=4$.  The geometry dependent
term $[F(\theta_1,\eta_{p,\rm{max}})]^{1/2}$ is plotted in
Fig.~1a as the dashed curves for different $\eta_{p,\rm{max}}$
values and $r_c=4$.  It can be seen that highly oblique shocks
allow proton acceleration on very short time scales. In the
limiting case of a perpendicular shock, drift--shock acceleration
drives the energy gain, and the finite size of the shock front
restricts the maximum particle energy.  The maximum proton
Lorentz factor $\gamma_p$ is reached for the maximum drift
distance, the shock size, and is given by
\begin{equation}
\gamma_{p,\rm{max}} = 3.2 \times 10^{-7} \beta_1 R B
\end{equation}
with $R$ in cm and $B$ in Gauss.

At their maximum Lorentz factors, the acceleration process for
protons is considerably slower than for electrons, and so the
proton acceleration time must be consistent with the observed
variability time, $t_{\rm var}$,
\begin{equation}
t_{\rm var} D \geq t_{\rm{acc},p,\rm{max}}
\label{eq:varcon}
\end{equation}
where $D$ the Doppler factor.  This can be converted to a
constraint on the geometry dependent term using
Eq.~\ref{eq:tacc_p}, and is plotted as a function of $\theta_1$
in Fig.~1a for a typical set of TeV-blazar parameters ($B\approx
20$~G, $D \approx 10$, $t_{\rm var}=12$~hours). The region below
the solid line is allowed by the variability constraint
(Eq.~\ref{eq:varcon}), and gives for each $\eta_{p,\rm{max}}$
value a minimum shock angle between $69^\circ$ and $88^\circ$,
depending on $\eta_{p,\rm{max}}$.  Thus, proton shock
acceleration on hour time scale in hadronic models can only take
place in oblique shocks, and the maximum drift distance may
restrict the maximum energy gain rather than the gyroradius.

Using the same shock geometry and magnetic turbulence spectra for
both protons and electrons, we find that the ratio
$F(\theta_1,\eta_{e,\rm{max}})/F(\theta_1,\eta_{p,\rm{max}})$
does not vary by more than a factor of 2 within the allowed shock
angle range (see Fig.~1b), allowing us to adopt an average value
for this ratio for a given parameter combination.  Inserting this
ratio into Eq.~\ref{eq:gammamax_ratio} then restricts the ratio
of the allowed maximum particle energies to the range below the
solid lines shown in Fig.~2. Points exactly on this line
represent models where the particle spectra are limited by
synchrotron losses and the acceleration time scale is exactly the
variability time scale; points below this line apply if adiabatic
losses are dominant for protons or the proton acceleration time
scale is shorter than the variability time scale in the jet
frame.  We note that for a given maximum proton energy, the
highest maximum electron energy occurs with Bohm diffusion.

In hadronic blazar models pion photoproduction is essential for
neutrino production. The threshold for this process is given by
$\epsilon_{\rm{max}} \gamma_{p,\rm{max}} = 0.0745$~GeV where
$\epsilon_{\rm{max}}$ is the maximum photon energy of the target
field, which in the Synchrotron Proton Blazar models is in turn
produced by the co-accelerated $e^-$.  In the $\delta$-function
approximation for the synchrotron emission, $\epsilon_{\rm{max}}
= (3/8) \gamma_{e,\rm{max}}^2 (B/B_{\rm{cr}}) m_e c^2$ with
$B_{\rm{cr}} = 4.414 \times 10^{13}$~G.  Inserting
$\epsilon_{\rm{max}}$ into the threshold condition, we find
\begin{equation}
\gamma_{p,\rm{max}} \ge 1.72 \times 10^{16} \left(\frac{B}{\rm{1
\; Gauss}}\right)^{-1} \gamma_{e,\rm{max}}^{-2}
\end{equation}
and this is shown in Fig.~2 as the dashed lines for various
magnetic field strengths.  Together with Eq.~11 the allowed range
of maximum particle energies is then restricted to the area below
the solid lines and above the dashed lines in Fig.~2, as shown
for example as the shaded area for $B\approx 20$~G and Kolmogorov
turbulence.  Inspecting hadronic models as presented in the
literature (e.g. Mannheim 1993, Mannheim et al 1996, Rachen
1999), we find that most models which are able to fit the
observations lie above the Bohm diffusion line (see Fig.~2),
indicating that the turbulence spectrum required in common
hadronic blazar jet models is likely to be of
Kolmogorov/Kraichnan type.
Rachen (1999) speculates that the transition between Kolmogorov and Kraichnan type
turbulence could be
responsible for the difference between low- and high frequency peaked
BL~Lacs.

\begin{figure}[b!] 
\centerline{\epsfig{file=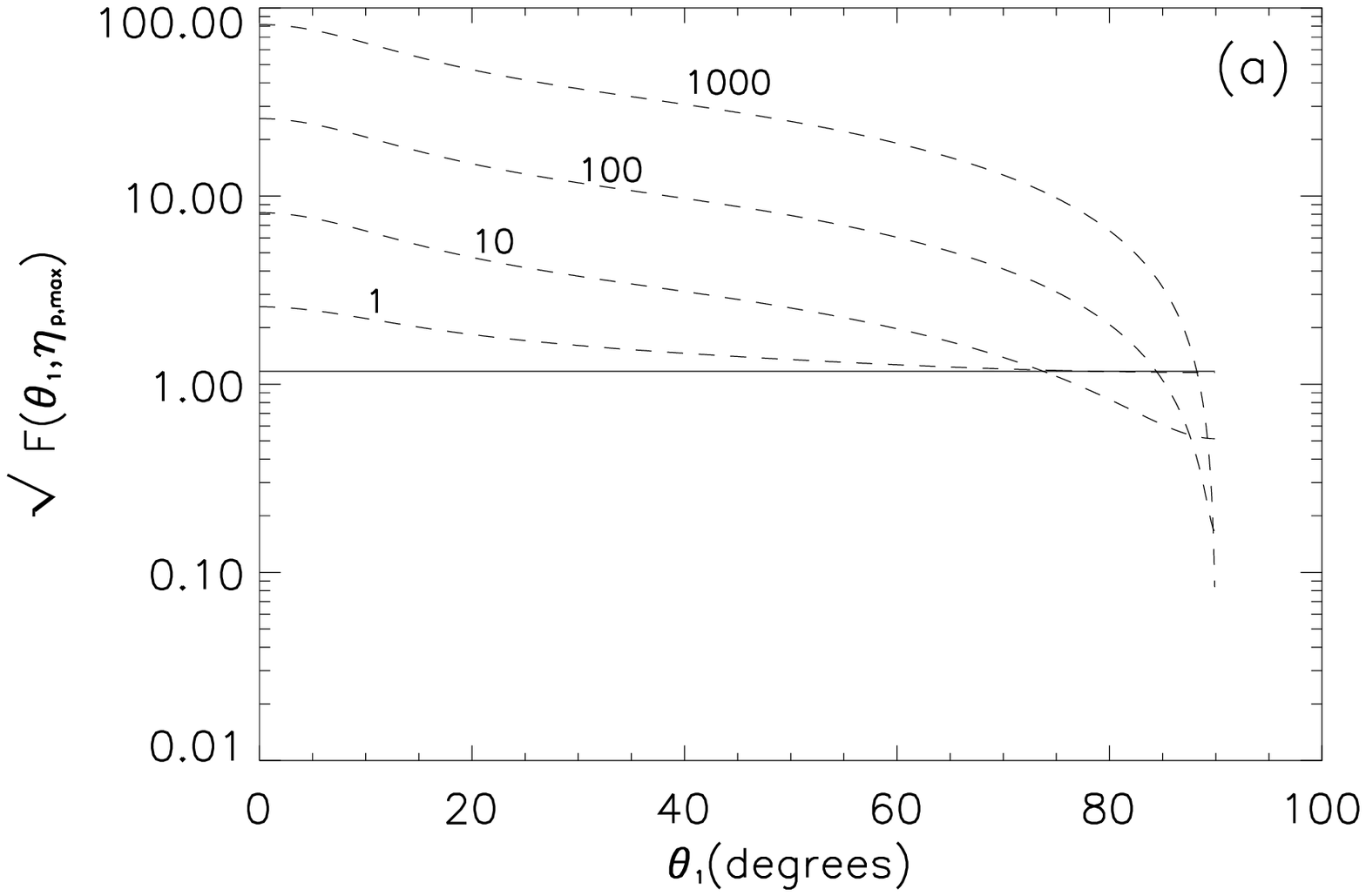,height=3.5in,width=3.5in}}
\vfill
\centerline{\epsfig{file=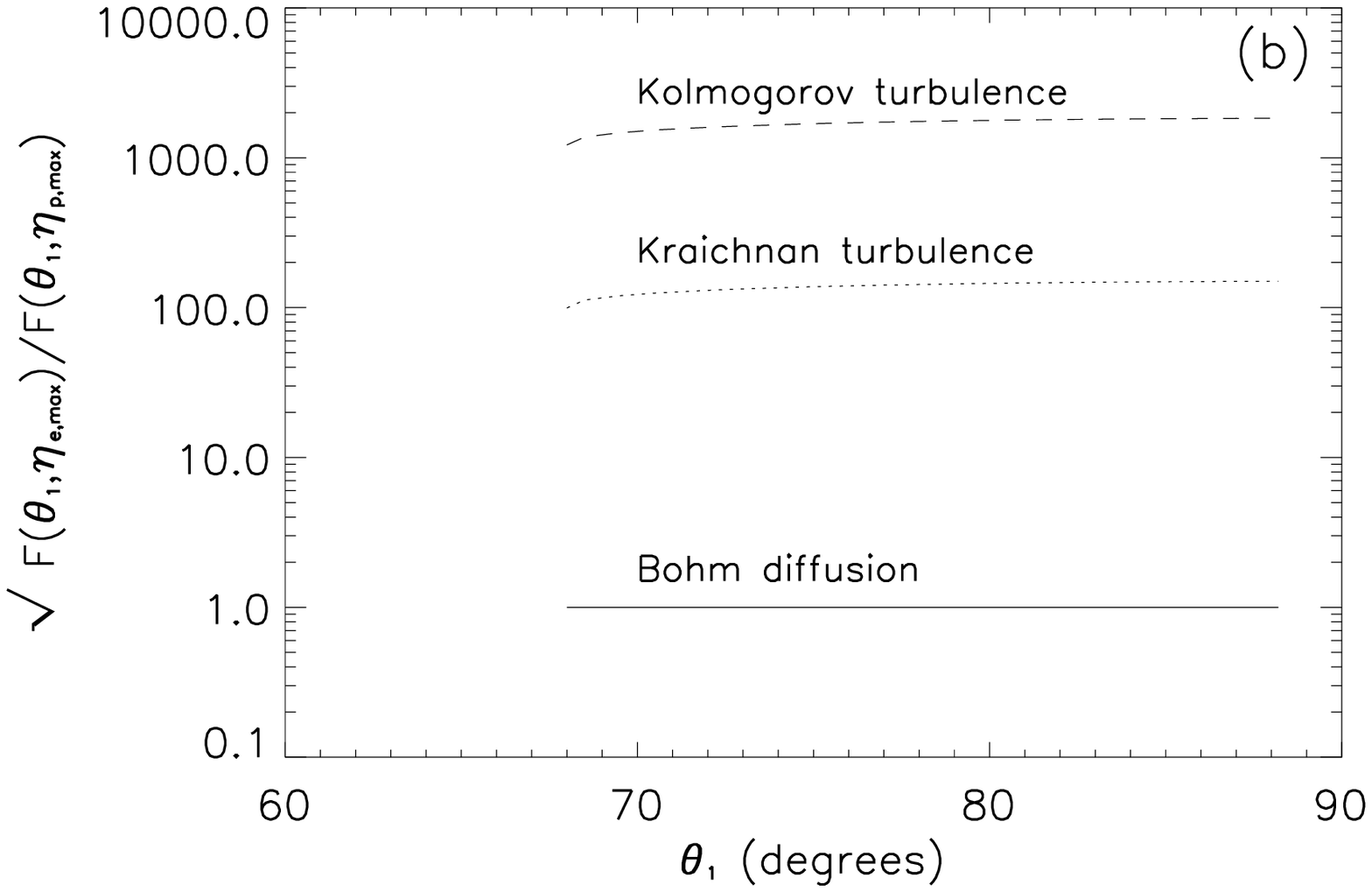,height=3.5in,width=3.5in}}
\vspace{10pt}
\caption{(a) $\sqrt{F(\theta_1,\eta_{p,\rm{max}})}$ versus shock
angle (dashed lines) for different $\eta_p$ values (numbers
attached to the curves) and $r_c=4$. Note that the diffusion
approximation limits the mean free path to $\eta <
\beta/\beta_1$. Region below the solid line satisfies
$t_{\rm{var}} D \geq t_{acc,p,\rm{max}}$ for a typical set of
TeV-blazar parameters $B\approx 20$~G, $D \approx 10$,
$u_1=0.5c$, $\beta=1$, $r_c = 4$, $t_{var}=12$~hours. (b) Ratio
$\sqrt{F(\theta_1,\eta_{e,\rm{max}})/F(\theta_1,\eta_{p,\rm{max}})}$
versus $\theta_1$ for different turbulence spectra.}
\label{fig1}
\end{figure}

\begin{figure}[b!] 
\centerline{\epsfig{file=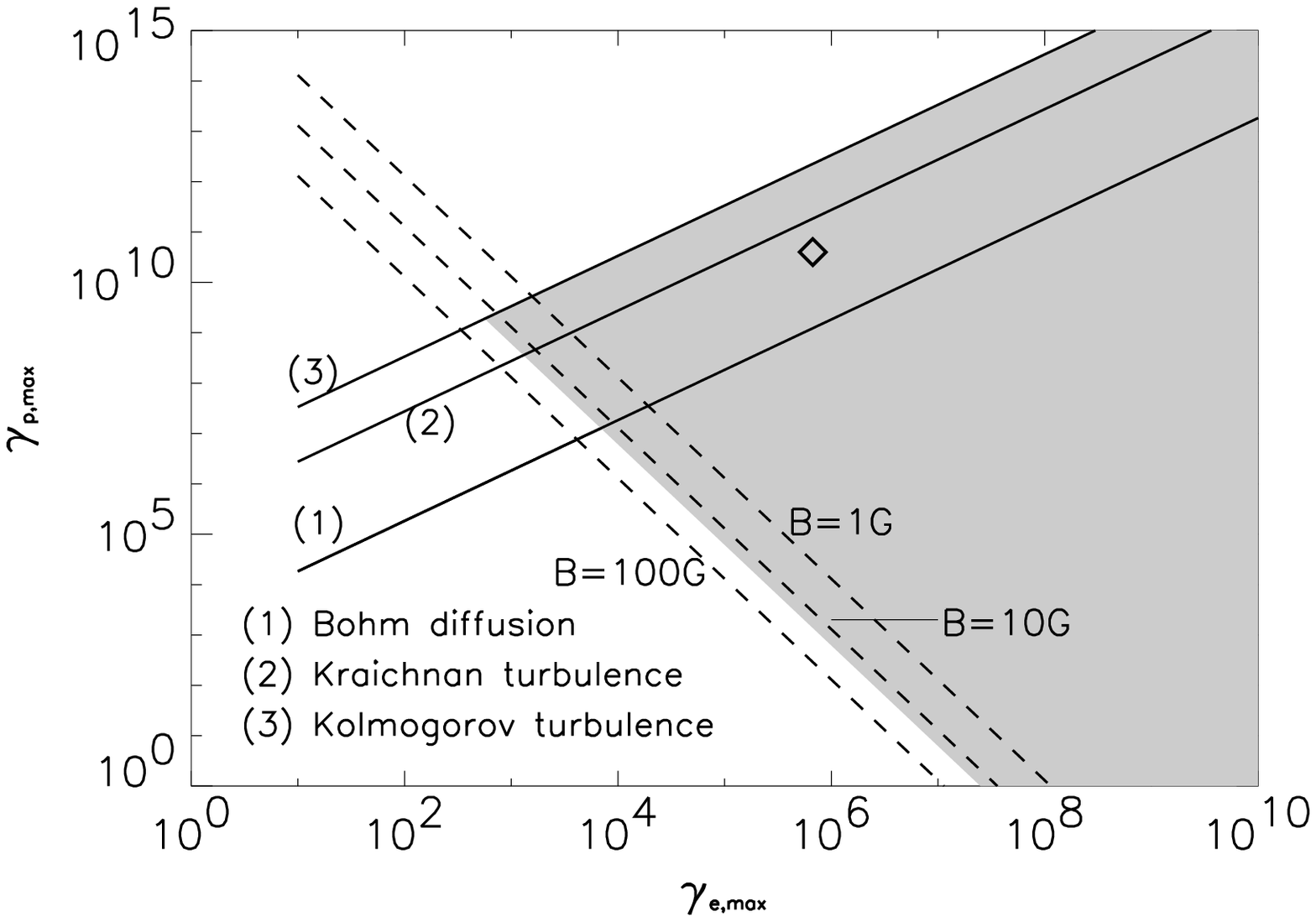,height=3.5in,width=3.5in}}
\vspace{10pt}
\caption{Allowed parameter space (shaded area) for
$\gamma_{p,\rm{max}}$, $\gamma_{e,\rm{max}}$ in the SPB-model for
different magnetic turbulence spetra. The diamond symbol
corresponds to the Mkn~501-model presented below ($B\approx
20$~G, $D \approx 10$, $u_1=0.5c$, $\beta=1$,
$t_{\rm{var}}=12$~hours).}
\label{fig2}
\end{figure}


\section{Emission Processes}

In the present model, the co-accelerated $e^-$ are assumed to
produce most of the {\it{observed}} low energy part of the
SED by synchrotron radiation, 
and this is assumed to be the target
radiation field for $p\gamma$ interactions and subsequent
cascading.  The observed hardening of the spectrum with rising
flux, has recently been convincingly reproduced by a shock model
with escape and synchrotron losses (Kirk et al 1998).  The
spectral slopes in this model are controlled by synchrotron
cooling, and thus naturally explain the temporal behaviour of the
spectral index as observed in the X-ray band.  The flaring
behaviour is explained by the shock front running into plasma
whose density is locally enhanced, and which thus increases the
number of particles injected into the acceleration process. The
increase in the plasma density is accompanied by an increase in
the magnetic field, leading to a higher acceleration rate and a
shift of the maximum particle energies to higher energies. This
picture departs from the standard explanation in which the flat
synchrotron spectra appear as a result of a superposition of
several local self-absorbed synchrotron spectra with changing
self-absorption frequency, adopted in previous PIC models
(e.g. Mannheim 1993).  

For simplicity, and because we do not wish to include additional
parameters, we use the same magnetic field for synchrotron
radiation as for acceleration.  For normal shocks this
approximation is justified as the magnetic fields either side of
the shock are similar.  This approximation might even be
justified for oblique shocks as a lower magnetic field in the
upstream region, compared to the downstream region, implies a
higher diffusion coefficient and time spent upstream, increasing
the synchrotron losses there and partially compensating for
having a lower field upstream.  Also, at oblique shocks,
reflection at the shock front itself is thought to be more
important than diffusion in the downstream region (Kirk \&
Heavens 1989), so that accelerating particles spend most of their
time upstream.  In addition, we assume that pitch-angle
scattering maintains quasi-isotropic particle distributions, and
all radiating particles are confined to the homogeneous emission
region.


\subsection{Energy Losses}

There are several energy loss/interaction processes which are
important for protons, electrons and photons in a dense radiation
field produced by relativistic electrons co-accelerated along
with the protons: protons interact with photons, resulting in
pion production and (Bethe-Heitler) pair production; electrons,
muons, protons and charged pions emit synchrotron radiation;
photons interact with photons by pair production.  We shall show
below that, for the present model, Inverse Compton emission by
the electrons can be neglected.

For simplicity we represent the observed synchrotron spectrum of
Mrk~501 during flaring, the target photon field for the
$p\gamma$-collisions and photon-photon pair production, as a
broken power-law:
\begin{equation}
n(\epsilon) \propto \left\{ \begin{array}{ll} \epsilon^{-1.6} &
 \mbox{for} \qquad 10^{-7}\rm{eV} \leq \epsilon \leq
 1.6~\rm{keV}\nonumber\\ \epsilon^{-1.8\phantom{/2}} & \mbox{for}
 \qquad 1.6~\rm{keV} \leq \epsilon \leq 42~\rm{keV}
\end{array} \right.
\label{eq:broken_pl}
\end{equation}

For determining the photon density of the target field, the
dimension of the emission region, assumed to be spherical, must
be known.  This can be estimated by setting the photon crossing
time equal to the variability time scale (in the jet frame -- in
the remainder of this section all quantities are in the jet frame
unless noted otherwise), making the implicit assumption that the
light crossing time scale determines the flux variations.  The
observed variability time scale can, in general, depend on: (i)
the injection time scale for the energetic particles
$t_{\rm{acc}}(E)$; (ii) the time needed for converting their
energy into radiation, i.e. their energy loss time scale
$t_{\rm{loss}}(E)$; (iii) the effective light crossing time
$t_{\rm{cross}}(E)\approx (2R_{\rm{blob}}/c) \times
P_{\rm{esc}}(E)$ where $R_{\rm{blob}}$ is the geometrical blob
radius and $P_{\rm{esc}}(E)$ is the energy dependent probability
for photons escaping from the blob taking account of
$\gamma\gamma$-pair production and diffusion during cascading.
Hence,
\begin{equation}
D t_{\rm{var}} \approx \max(t_{\rm acc}, t_{\rm{loss}},
t_{\rm{cross}}).
\end{equation}

For leptonic models, the time scales for energy losses and
acceleration are typically significantly shorter than the
crossing time, i.e. $t_{\rm{loss}}$,$t_{\rm acc} \ll
t_{\rm{cross}}$, and thus the radius of the emission region can
be derived from the observed variability time scale. In addition,
the emission region is assumed to be optically thin at 1 TeV and
X-ray energies, implying that $R_{\rm{TeV}} \approx R_{\rm{X}}$,
where $R_{\rm{TeV}}$, $R_{\rm{X}}$ are the dimensions of the
emitting region at 1~TeV and X-ray energies, respectively.  This
differs in two points from the hadronic blazar jet models:
Firstly, the optical depth of the emission region is strongly
energy dependent, leading to an effective, i.e. observed,
thickness of the emission region $R_{\rm{eff}}(E) \approx
R_{\rm{blob}} P_{\rm{esc}}(E)$, which also depends on the energy.
Diffusion can be neglected during cascading since the cascade
processes are of leptonic origin, and are, in general, more rapid
than diffusion.  Thus, in SPB models the crossing time scale in
the optically thick TeV-band is related to the crossing time
scale at X-ray energies (optically thin), through
\begin{equation}
t_{\rm{cross,TeV}} \approx
t_{\rm{cross,X}}[1-\exp(-\tau_{\gamma\gamma,\rm{TeV}})]
/\tau_{\gamma\gamma,\rm{TeV}}
\end{equation}
(averaging over a homogeneous emission volume; see Rachen 1999).
Hence, $t_{\rm{cross,X}}$ is the relevant time scale for
estimating the radius $R_{\rm{blob}}\approx R_X$ of the emission
region.  Secondly, the acceleration and/or energy loss time
scales can be of the same order of magnitude as the crossing time
scale. Acceleration, however, is always faster than the energy
losses of the accelerating particles up to their maximum energy.
Note that because of the leptonic nature of the cascade
processes, the energy loss time scale will in general be
determined by the (slower) hadronic processes. As a further
consequence, flux variations are not significantly washed out by
the cascading mechanism, but closely follow the crossing or
hadronic loss time scales.

If $t_{\rm{loss}} \leq t_{\rm{cross}}$ the crossing time scale
determines the flux variations, and in this case we can estimate
the radius of the emission region through $R_{\rm{blob}} \approx
0.5 c D t_{\rm{var,X}}$ with $t_{\rm{var,TeV}} \approx
t_{\rm{var,X}}/\tau_{\gamma\gamma,\rm{TeV}}$.  In our present
model, efficient proton synchrotron emission determines the
TeV-bump in the blazar SED, and so at the maximum proton energy
$t_{\rm{ad}} > t_{p, \rm{syn}}$.  The adiabatic loss time scale
due to expansion is (Longair 1994)
\begin{equation}
t_{\rm{ad}} = R_{\rm{ad}}^{-1} \equiv |R_{\rm{blob}}/\dot
R_{\rm{blob}}| = 2|B/\dot B| \approx R/u_1
\end{equation}
assuming magnetic flux conservation $B \propto R_{\rm
{blob}}^{-2}$, and so $t_{\rm{ad}}$ is related to the size of the
emission region $R_{\rm{blob}}$. Since $t_{\rm{loss}} \approx
t_{p, \rm{syn}}$ for the highest proton energies in our model
responsible for the TeV-emission, $t_{\rm{loss}} \leq
t_{\rm{cross}}$, and so the size of the emission region can be
determined by the variability time scale.

The shortest doubling time measured by the Whipple Telescope in
the 1997 data of Mkn~501 was approximately 2 hours (Quinn et al
1999), while HEGRA reported a lower doubling time of 15 hours
(Aharonian et al 1999) or 12 hours (Krawczynski 1999).  As a
working hypothesis we adopt here a variability time scale of
12~hours, but will discuss also effects of smaller variability
time scales.  For $t_{\rm{var}} \approx 12$~hours we find
$R_{\rm{blob}}\approx 8\times 10^{15}$cm for $D \approx 10$.

Eq.~12 and 13 imply that the acceleration time scale for electrons
is much smaller than for protons at their
maximum energies, $t_{acc,e,\rm{max}} \ll t_{acc,p,\rm{max}}$. If
a $\gamma$-ray outburst thus corresponds to particle acceleration
of a single $p/e^-$ population, then an increase of the low energy
synchrotron flux will occur before the TeV-flare. The amount of the lag 
depends on the proton acceleration time, and so on the shock parameters, in particular the diffusion
coefficients. E.g. 
for the present model and the Kolmogorov spectrum $t_{acc,p,\rm{max}}
\approx 10^5$~s (see Fig.~3), which for D=10 suggests a time lag of
$\approx 5$~hours. For Bohm diffusion, the time lag would be $\approx 14$~hours.
These time lags are consistent with current observational constraints.

The relevant radiation and loss time scales for photomeson
production, Bethe-Heitler pair production, $p$ synchrotron
radiation, and adiabatic losses due to jet expansion, are shown
in Fig.~3 together with the acceleration time scale.  Synchrotron
losses, which turn out to be at least as important as losses due
to photopion production in our model, limit the injected $p$
spectrum to a Lorentz factor of $\gamma_p \approx 3 \times
10^{10}$ for the assumed model parameters.  We adopt a Kolmogorov
spectrum of turbulence for the magnetic field structure
($\delta=5/3$), and so for any $\eta_p \ge 1$ value, variability
arguments constrain the shock angle to $\theta_1 \geq 75^\circ$
(see Fig.~1). The maximum proton energy could then be achieved,
e.g., with $\eta_p = 10$, $\theta_1 =85^\circ$ and $u_1=0.5c$.
This is in agreement with the limit imposed on
quasi-perpendicular shocks due to their finite shock size.  Note
that due to the non-zero shock angle, the acceleration time scale
shown in Fig.~3 does not follow a strict power-law, but is
curved. This is due to the non-linear dependence of
$F(\theta_1,\eta_{p})$ on the particle's gyroradius (see
Eqs.~6--8).  A $\gamma_p^{-2}$ proton spectrum, typical of shock
accelerated particles, is used for $2\leq\gamma_p\leq \gamma_{p,
{\rm max}}$, where $\gamma_{p, {\rm max}} = 3\times 10^{10}$ is
obtained from requiring $t_{{\rm acc,}p}=t_{{\rm syn,}p}$ at
$\gamma_{p, {\rm max}}$.

Rachen \& M\'esz\'aros (1998) noted the importance of synchrotron
losses of $\mu^\pm$ (and $\pi^\pm$) prior to their decay in AGN
jets and GRBs.  The critical Lorentz factors $\gamma_{\mu}
\approx 2 \times 10^9$ and $\gamma_{\pi} \approx 4 \times
10^{10}$, above which synchrotron losses in the assumed magnetic
field dominate over decay, lie below the maximum Lorentz factor
for $\mu^\pm$ and above the maximum Lorentz factor for
$\pi^\pm$. Thus, while $\pi^\pm$-synchrotron losses can safely be
ignored, $\mu^\pm$-synchrotron losses should be included.

For the parameters employed in this work we receive 
a target photon energy density
of $u_{\rm{target}} \approx 0.06~\rm{TeV/cm}^3$, and a magnetic
field energy density of $u_B \approx 11.7~\rm{TeV/cm}^3$ using
$B\approx 20$~G. With
$u_B \gg u_{\rm{target}}$ significant Inverse Compton radiation
from the co-accelerated $e^-$ is not expected.

We also neglect secondary production interactions of relativistic
protons with the ambient thermal plasma. In the present model the
dominant loss process turns out to be proton synchrotron radiation. Thus,
our assumption is justified if the thermal proton density
does not exceed $\approx 10^{9\ldots 10}$~cm$^{-3}$.
In Section~4 we estimate the number density of cold protons to be less than this
for reasonable values of the jet width.

\begin{figure}[h] 
\centerline{\epsfig{file=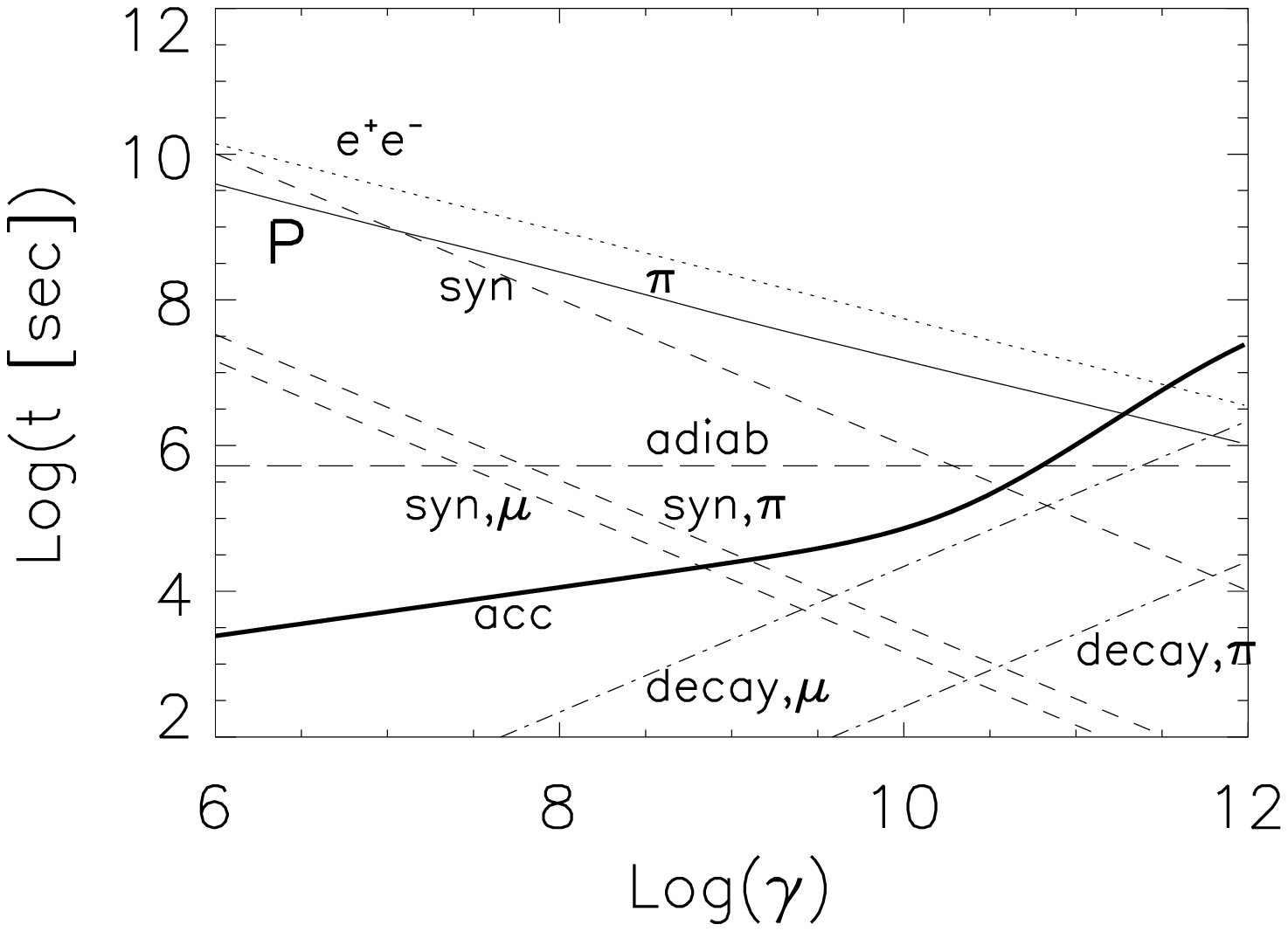,height=3.5in,width=3.7in}}
\vspace{10pt}
\caption{ Mean energy loss time of $p$ for $\pi$-photoproduction
($\pi$), Bethe-Heitler pair production ($e^+e^-$) and synchrotron
radiation (syn).  Loss times for $\pi^\pm$- and $\mu^\pm$ for
synchrotron radiation (syn $\pi$, syn $\mu$) are also shown and
compared with their mean decay time scales (decay $\pi$, decay
$\mu$). The acceleration time scale (acc), based on Kolmogorov
turbulence, is calculated for $r_c = 4$, $u_1 = 0.5c$ and shock
angle $\theta_1 = 85^\circ$. The curvature of the acceleration
time scale is caused by the non-linear dependence of
$F(\theta,\eta_{p})$ on $\gamma$.  The adiabatic loss time
(adiab) is assumed to be $2|B/\dot B| \approx R/u_1 \approx D
t_{\rm{var}}$.  We adopt $B\approx 20$~G, and all quantities are
in the co-moving frame of the jet.}
\label{fig2}
\end{figure}


\subsection{Simulation of particle production and cascade development}

For the first time in the context of the SPB-model, we use the
Monte-Carlo technique to simulate particle production and cascade
development, and this allows us to use exact cross sections.  For
photomeson production we use the Monte-Carlo code SOPHIA (M\"ucke
et al 2000), and Bethe-Heitler pair production is simulated using
the code of Protheroe \& Johnson (1996). We calculate the yields
for both processes separately, and the results are then combined
according to their relative interaction rates.

The mean pion production interaction rate for an isotropic photon
field is
\begin{equation}
r_{\pi}(E_p) = \frac{1}{8E_p^2\beta_p} \int_{\epsilon_{\rm
th}}^{\infty}\!d\epsilon\, \frac{n(\epsilon)}{\epsilon^2}
\int_{s_{p,\rm th}}^{s_{p,\rm max}}\!ds_p\,(s_p-m_p^2)
\sigma_{\pi}(s_p)\ ,
\end{equation}
where $s_p = m_p^2 + 2E_p\epsilon (1-\beta_p\cos\theta_p)$ is the
center-of-momentum (CM) energy squared, $\theta_p$ the angle
between the proton and the photon, $\beta_p c$ the proton's
velocity, $\sqrt{s_{p,\rm th}} \approx 1.08$GeV the threshold CM
energy, $\epsilon_{\rm th} = (s_{p,\rm{th}}-m_p^2)/2 (E_p+p_p)$,
$s_{p,\rm max} = m_p^2+ 2E_p\epsilon(1+\beta_p)$ and
$\sigma_{\pi}$ the pion production cross section.  The
Bethe-Heitler pair production interaction rate, $R_{\rm{BH}}$, is
calculated using the formulae given in Chodorowski (1992).

In highly magnetized environments, proton-photon interactions
compete with synchrotron radiation by the protons. To take proton
synchrotron losses into account in our code, we sample a
$p\gamma$ interaction length from an exponential distribution
with its corresponding mean $\bar x_{p\gamma} = c/r_{\pi,BH}$ as
given below.  We make the approximation that the proton energy on
interacting is given by
\begin{equation}
E_p \approx E_p^{(0)} \left[1 + {x_{p \gamma} \over
\bar{x}_{p,syn}(E_p^{(0)})} \right]^{-1}
\end{equation}
where $E^{(0)}_{p}$ is the initial proton energy, $x_{p\gamma}$
is the sampled interaction length for pion production or
Bethe-Heitler pair production, and $\bar{x}_{p,\rm{syn}} =
c/r_{p,\rm{syn}}$ is the proton synchrotron loss distance given
by
\begin{equation}
r_{M,\rm{syn}} = \frac{4}{3} \left(\frac{m_e}{M}\right)^2
{\sigma_T \gamma u_B \over M c}.
\label{eq:synrate}
\end{equation}
where $M=m_p$, $\sigma_T$ is the Thomson cross section, and $u_B$
is the magnetic energy density.

The relatively long mean life time of highly energetic charged
pions, muons and kaons might be of the same order of magnitude as
their synchrotron loss time scale in highly magnetized
environments like AGN jets and GRBs (Rachen \& M\'esz\'aros 1998). We
simulate their synchrotron energy losses by sampling the decay
length $x_{\rm{dec}}$ from an exponential distribution with
corresponding mean decay length $\bar x_{\rm{dec}} = c\gamma
\tau_{\rm{dec}}$. For $\mu^\pm$ we have $\tau_{\rm{dec}} =2.20
\times 10^{-6}$~s, while $\tau_{\rm{dec}} = 2.60 \times
10^{-8}$~s for $\pi^\pm$ and $\tau_{\rm{dec}} = 1.24 \times
10^{-8}$~s for $K^\pm$.  The particle's energy on decaying is
then
\begin{equation}
E_{K,\pi,\mu} = E^{(0)}_{K,\pi,\mu} \left[1 + {2 x_{dec} \over
\bar{x}_{K,\pi,\mu,\rm{syn}}(E^{(0)}_{K,\pi,\mu})} \right]^{-1/2}.
\end{equation}
where $E^{(0)}_{K,\pi,\mu}$ is the initial $K$,$\pi$, or $\mu$
energy, $\bar x_{K,\pi,\mu,\rm{syn}} = c/r_{K,\pi,\mu,\rm{syn}}$
is the synchrotron loss distance with $r_{K,\pi,\mu,\rm{syn}}$ 
given by Eq.~\ref{eq:synrate} for $M=m_K$, $m_\pi$ or $m_\mu$.

We next outline the simulations of the cascade development in
more detail.  Energetic photons will pair produce on the target
photon field, and if the magnetic energy density exceeds the
injected target field density ($u_{\rm{B}} > u_{\rm{target}}$)
they will initiate a pair-synchrotron cascade.  Energetic photons
arise directly from $\pi^0$-decay, or indirectly as synchrotron
photons from protons, charged mesons and electrons resulting from
$\pi^\pm\rightarrow \mu^\pm\rightarrow e^\pm$ decay and
Bethe-Heitler pair production.  The optical depth
$\tau_{\gamma\gamma}(E_{\gamma})$ for $\gamma$-ray photons with
energy $E_{\gamma}$ for $e^\pm$ pair production inside the blob
is given by
\begin{equation}
\tau_{\gamma\gamma}(E_{\gamma}) =
\frac{R_{\rm{blob}}}{8E^2_\gamma}
\int_{\epsilon_{\rm{min}}}^{\infty} d\epsilon
\frac{n(\epsilon)}{\epsilon^2}
\int_{s_{\gamma,\rm{min}}}^{s_{\gamma,\rm{max}}(\epsilon,E_{\gamma})}
ds_\gamma s_\gamma \sigma_{\gamma\gamma}(s_\gamma)
\end{equation}
where $n(\epsilon)$ is the differential photon number density and
$\sigma_{\gamma\gamma}(s)$ is the total cross section for
photon-photon pair production (Jauch \& Rohrlich 1955) for a
centre of momentum frame energy squared given by
\begin{equation}
s_{\gamma} = 2\epsilon E_\gamma(1-\cos{\theta_{\gamma}})
\end{equation}
where $\theta_{\gamma}$ is the angle between directions of the
energetic photon and soft photon, and $s_{\gamma,\rm{min}} = (2
m_e c^2)^2$, $s_{\gamma,\rm{max}} = 4\epsilon E_\gamma$ and
$\epsilon_{\rm{min}} = (m_e c^2)^2/E_\gamma$.  For simulating
photon-photon pair production we approximate
$E_{e^+}=E_{e^-}=E_\gamma/2$. The radiating $e^\pm$ are assumed
to be continuously isotropized in the blob frame by deflection in
the uniform magnetic field, resulting in the synchrotron
radiation being isotropic in the blob frame.  The spectrum of
synchrotron photons, averaged over pitch-angle because of the
isotropy of the particle distribution, is calculated using
functions given by Protheroe (1990).

In general, the cascade can be initiated by photons from
$\pi^0$-decay (``$\pi^0$ cascade''), electrons from the
$\pi^\pm\to \mu^\pm\to e^\pm$ decay (``$\pi^\pm$ cascade''),
$e^\pm$ from the proton-photon Bethe-Heitler pair production
(``Bethe-Heitler cascade'') and $p$ and $\mu$-synchrotron photons
(``$p$-synchrotron cascade'' and ``$\mu^\pm$-synchrotron
cascade'').  Here, we assume the cascades develop linearly, and
this requires the photon field produced by the cascade, to be
negligible as a target field in comparison with the injected
synchrotron radiation due to co-accelerated $e^-$.  This
requirement can be expressed as $\tau_{\gamma\gamma,\rm{cas}} \ll
\tau_{\gamma\gamma,\rm{target}}$ with
$\tau_{\gamma\gamma,\rm{target}}$,$\tau_{\gamma\gamma,\rm{cas}}$
being the pair production optical depths of photons on the target
and cascade photon field, respectively. Our Monte-Carlo results
show this condition to be met for the present input.  To simplify
the calculation, the electrons are completely cooled instantly by
synchrotron radiation before pair production by the synchrotron
photons takes place.  This approximation is equivalent to
assuming that $t_{\rm{syn}} \ll t_{\rm{pair}}$ which is justified
because of the very short synchrotron life time of electrons in
the assumed magnetic field.

A matrix method (e.g. Johnson et al 1996) is then used to follow
the pair-synchrotron cascade in the ambient synchrotron radiation
field and magnetic field. The cascades are considered in the jet
frame. Here, electron and photon fluxes are represented by
vectors $G_j^k$ and $F_i^k$, which give the total number of
electrons in the energy bin at energy $E_j$ and number of photons
at energy $E_i$, respectively, in the $k$th cascade generation.
We use a logarithmic stepsize of 0.1 ranging from
$\log({E_e/1\,\rm{GeV}}) = -3$ to $12$ in electron energy, and
from $\log({E_\gamma/1\,\rm{GeV}}) = -13$ to $12$ in photon
energy.  Averaged over a homogeneous emission region, the
probability of gamma-ray interaction by photon-photon pair
production at energy $E_i$ is given by the vector
$P_{\rm{\gamma\gamma},i}$
\begin{equation}
P_{\rm{\gamma\gamma},i}=[1-P_{\rm{esc}}(E_i)] = \left[
1-\frac{1-\exp(-\tau_{\gamma\gamma}(E_{i}))}
{\tau_{\gamma\gamma}(E_{i})}\right].
\end{equation}

The transfer matrix $T^{(\rm{syn})}_{ij}$ gives the number of
synchrotron photons of energy $E_{i}$ produced by electrons of
energy $E_{j}$, $T^{(\rm{syn})}_{ij}$, and the transfer matrix
$T^{(\rm{pair})}_{ij}$ gives the number of $e^\pm$ of energy
$E_{j}$ produced through pair production of energetic photons of
energy $E_{i}$ on the target field.  The vectors and matrices are
calculated taking care of energy conservation.

The photon and electron fluxes due to synchrotron radiation and
photon-photon pair production are then calculated through matrix
multiplication. We start by calculating the number of
$\pi^0$-decay photons which pair produce in the blob 
$$
F_i^0 = I_i^0 P_{\rm{\gamma\gamma},i}
$$
where the vector $I_i^0$ gives the number of $\pi^0$-decay
gamma-rays at an energy $E_i$ (the emerging photons, i.e. those
which do not pair produce in the blob, are stored in an
array). The electron spectrum due to photon-photon pair
production is then given by
$$
F_j^1 = \sum_i F_i^0 T^{(\rm{pair})}_{ij}.
$$
These electrons radiate synchrotron photons, and the resulting
photon yield is determined by
$$
I_i^1 = \sum_j F_j^1 T^{(\rm{syn})}_{ij}.
$$
This is the 1st cascade generation photon spectrum, which in turn
again suffers photon-photon pair production on the target photon
field in the blob, etc. In our calculation, we iterate until
$F_i^k \leq 0.01$ at any energy $E_i$, or stop at 10 generations.

We have injected $10^4$ protons at each proton energy equally
spaced in $\log E_p$ at 0.1 decade intervals from $10^3$~GeV to
the maximum injection energy. These protons are assumed to be
continuously isotropized, and interact with the synchrotron
radiation field of the co-accelerated $e^-$ through pion
production and Bethe-Heitler pair production. The resulting
neutrinos (see Sect.~4.1) escape without further interaction, but
the $e^\pm$ and high energy $\gamma$-rays initiate cascades.  The
resulting particle spectra were weighted with $n_{p,0}E_p^{-2}
dE_p$, appropriate to the assumed proton injection spectrum, and
divided by the number of injected protons, and by $4\pi$
steradians.  Figs.~4 and 5 show examples of cascade spectra in
the observer's frame initiated by photons with different origins
for model parameters (given above), which satisfactorily
reproduce the flare spectrum of Mkn~501.

Our Monte-Carlo program only treats the first interactions of
protons, and so we must take account of this when making the
final flux predictions. In the case of pion production, a
fraction $\kappa \approx 0.25$ of the initial proton energy goes
into particle production, and $\sim 1/3$ of the time the emerging
nucleon is a neutron, and is assumed to escape from the blob
without further interaction.  Combining these two factors, the
resulting spectra must be multiplied by 2.01 to take account of
subsequent pion production interactions. Similarly, the
multiplication factor which takes account of subsequent
Bethe-Heitler pair production interactions is $\sim 10$.
Particle spectra due to pion production and Bethe-Heitler pair
production are then weighted according to their mean interaction
rates, i.e.  $R_{\pi}/R_{\rm{tot}}$ for pion production, and
$R_{\rm{BH}}/R_{\rm{tot}}$ for Bethe-Heitler pair production
where $R_{\rm{tot}} = R_{\pi} + R_{\rm{BH}} + R_{\rm{ad}}$ is the
total interaction rate, taking into account adiabatic losses of
the protons due to jet expansion approximated by Eq.~18.

Adiabatic jet expansion also affects the radius of the emission
region, its photon density and magnetic field, and thereby the
probability for $p\gamma$-interactions and the subsequent cascade
development. With the expansion of the emission region, the
probability for $p\gamma$-interactions decreases because of the
decrease of the photon energy density and magnetic field.  It
follows that $p\gamma$-interactions, and their resulting
cascades, most likely take place during the {\it{initial}} phase
of the jet expansion, and so we neglect any effects on the
emerging cascade spectra due to changes in magnetic field or
dimension of the emission region. In particular, the flare
rise-time can be much shorter than the expansion and light
crossing time scales if one associates it with the onset of the
$p\gamma$-interactions and their resulting cascades. Short
rise-times and longer flare decay time scales are typically
observed in blazar light curves.  Finally, we transform the
cascade spectrum to the observer's frame. For calculating the
distance, $q_0 = 0.5$ and $H_0 = 50$~km s$^{-1}$ Mpc$^{-1}$ are
used.

Cascades initiated by photons from $\pi^0$ decay, ``$\pi^0$
cascades'', or by electrons from $\pi\to\mu\to e$ decay,
``$\pi^\pm$ cascades'', (Fig.~4) produce rather featureless
spectra (see also Mannheim et al 1993).  However, cascades
initiated by protons ``$p$-synchrotron cascades'', and by muons from
$\pi\to\mu\to e$ decay, ``$\mu^\pm$-synchrotron cascades'' (Fig.~5b) produce
a double-humped SED as observed for $\gamma$-ray blazars (see
also Rachen 1999).  The contribution from Bethe-Heitler pair
production turns out to be negligible (Fig.~5a).  
In the model for Mkn~501 presented here, the synchrotron flux
from directly accelerated $e^-$ significantly exceeds
the reprocessed $\mu$ and $p$ synchrotron flux. Thus, in the
present Mkn~501 model direct proton
and muon synchrotron radiation is mainly responsible for the high
energy hump whereas the low energy hump is mainly synchrotron
radiation by the directly accelerated $e^-$.  In the present
model photons up to 1~TeV in the jet frame are optically thin to
$\gamma\gamma$-pair production. Thus, only a small fraction of the
power emitted as proton synchrotron radiation is redistributed to
lower energies via cascading. For models where the optical depth
of TeV-photons is significantly higher, a correspondingly larger
fraction of the TeV-bump may be redistributed to X-ray energies.
In these models, the pairs produced by photons of the ``high
energy hump'' may contribute considerably to the observed X-ray
bump, whereas in the present model they are negligible in
comparison to the directly accelerated electrons.  We note that
for the case of strong magnetic fields where
$\mu^\pm$-synchrotron cascades cause detectable humps, the $p$
cascade spectrum dominates in general over the $\mu^\pm$ cascade
spectrum.


\begin{figure}[h] 
\centerline{\epsfig{file=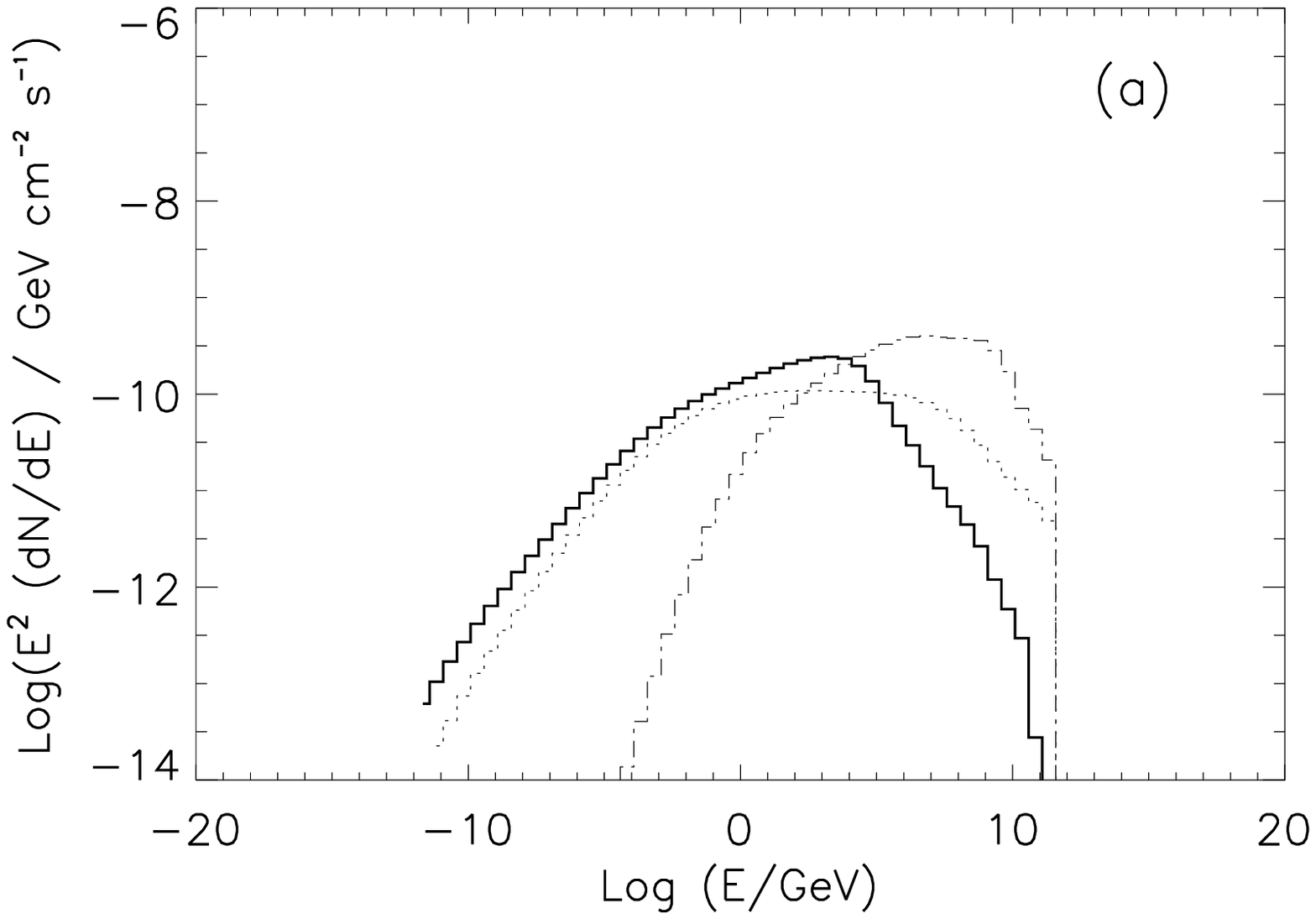,height=3.5in,width=3.7in}}
\hfill
\centerline{\epsfig{file=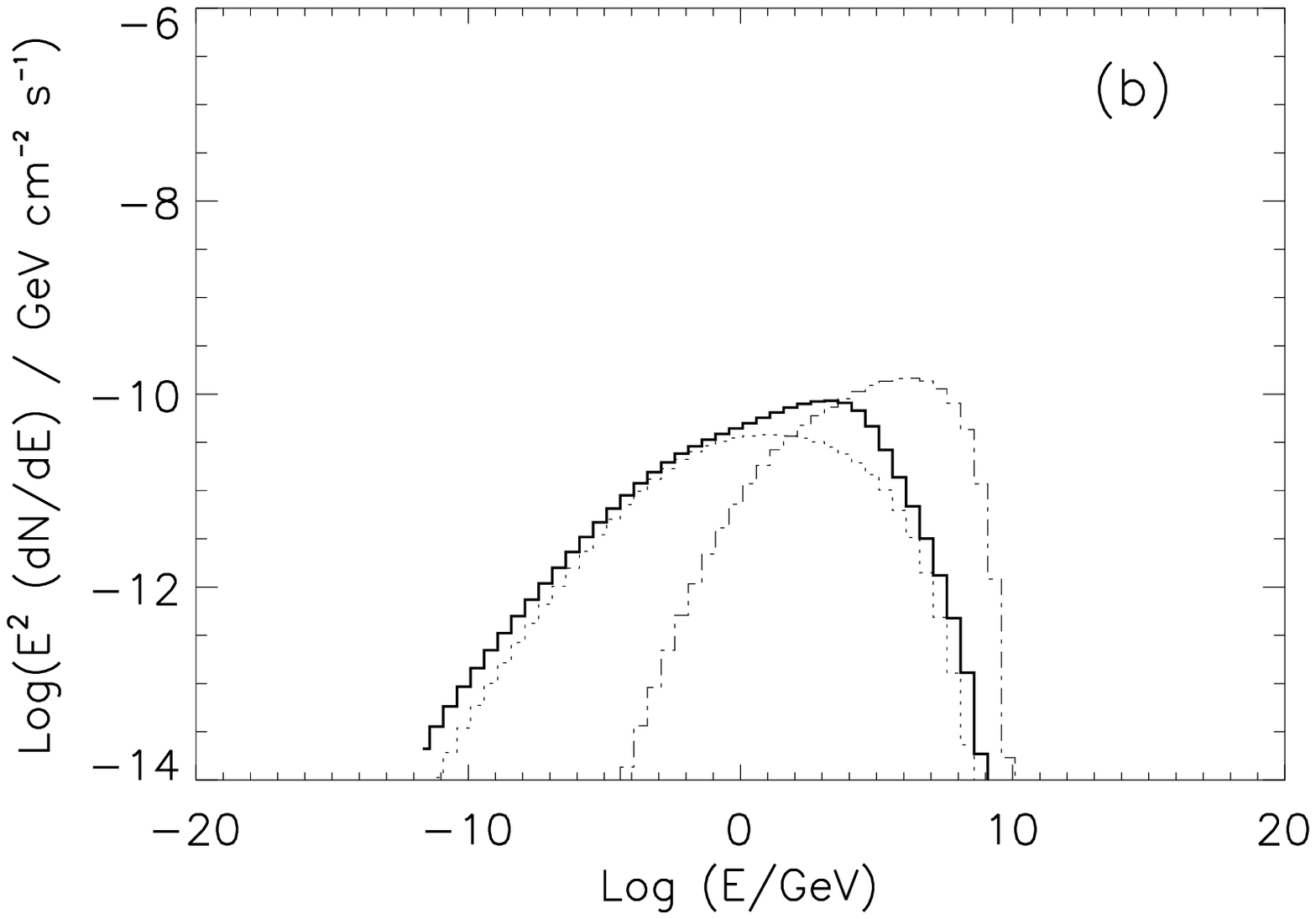,height=3.5in,width=3.7in}}
\vspace{10pt}
\caption{(a) Average cascade spectrum initiated by $\pi^0$-decay,
and (b) $\pi^\pm\rightarrow \mu^\pm\rightarrow e^\pm$-decay
synchrotron photons (lower panel). Dot-dashed and dotted
histogram shows the 1st and 2nd generation synchrotron photons on
production. Solid histogram shows the emerging photon spectrum on
production.}
\label{fig3}
\end{figure}

\begin{figure}[h] 
\centerline{\epsfig{file=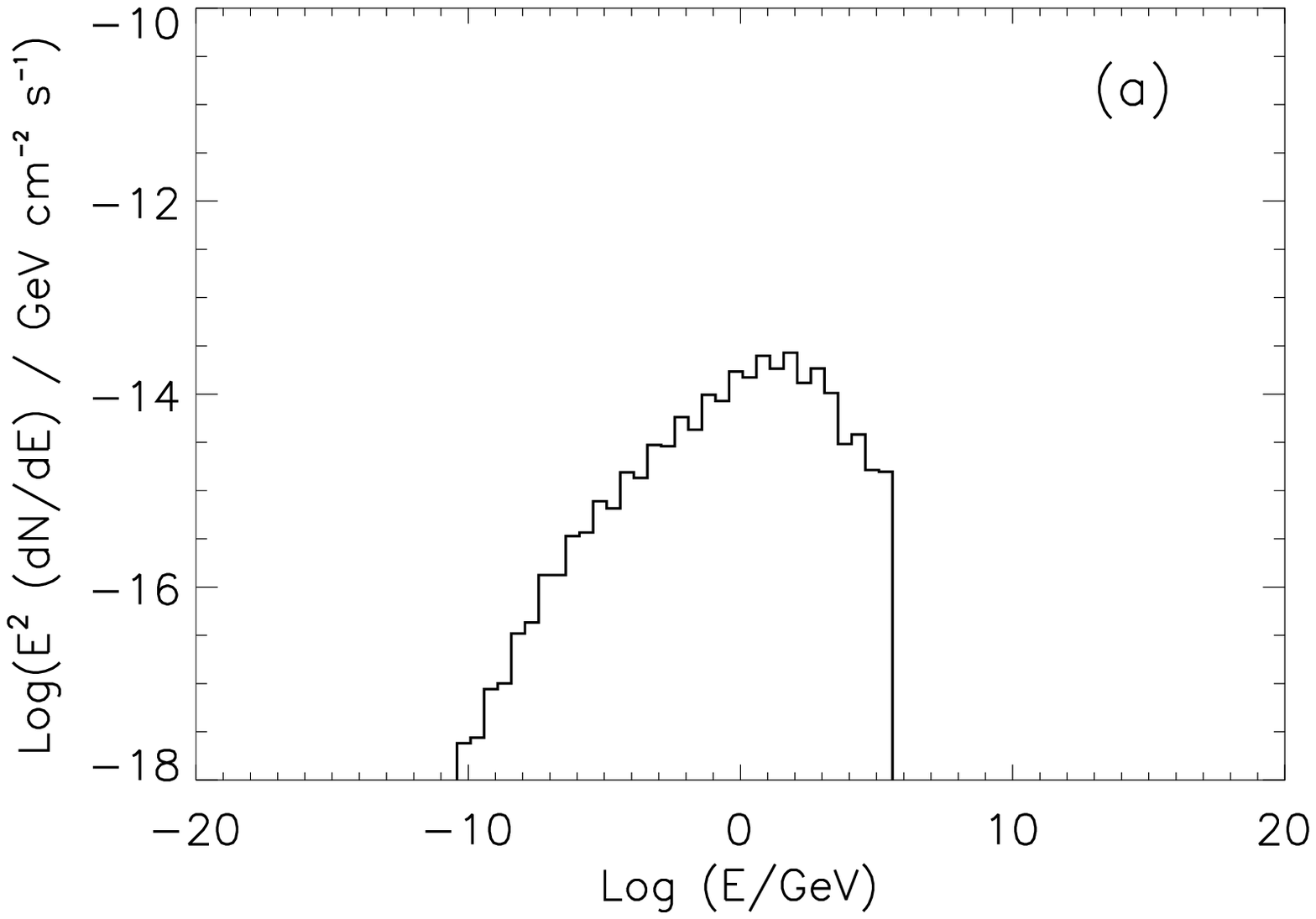,height=3.5in,width=3.7in}}
\hfill
\centerline{\epsfig{file=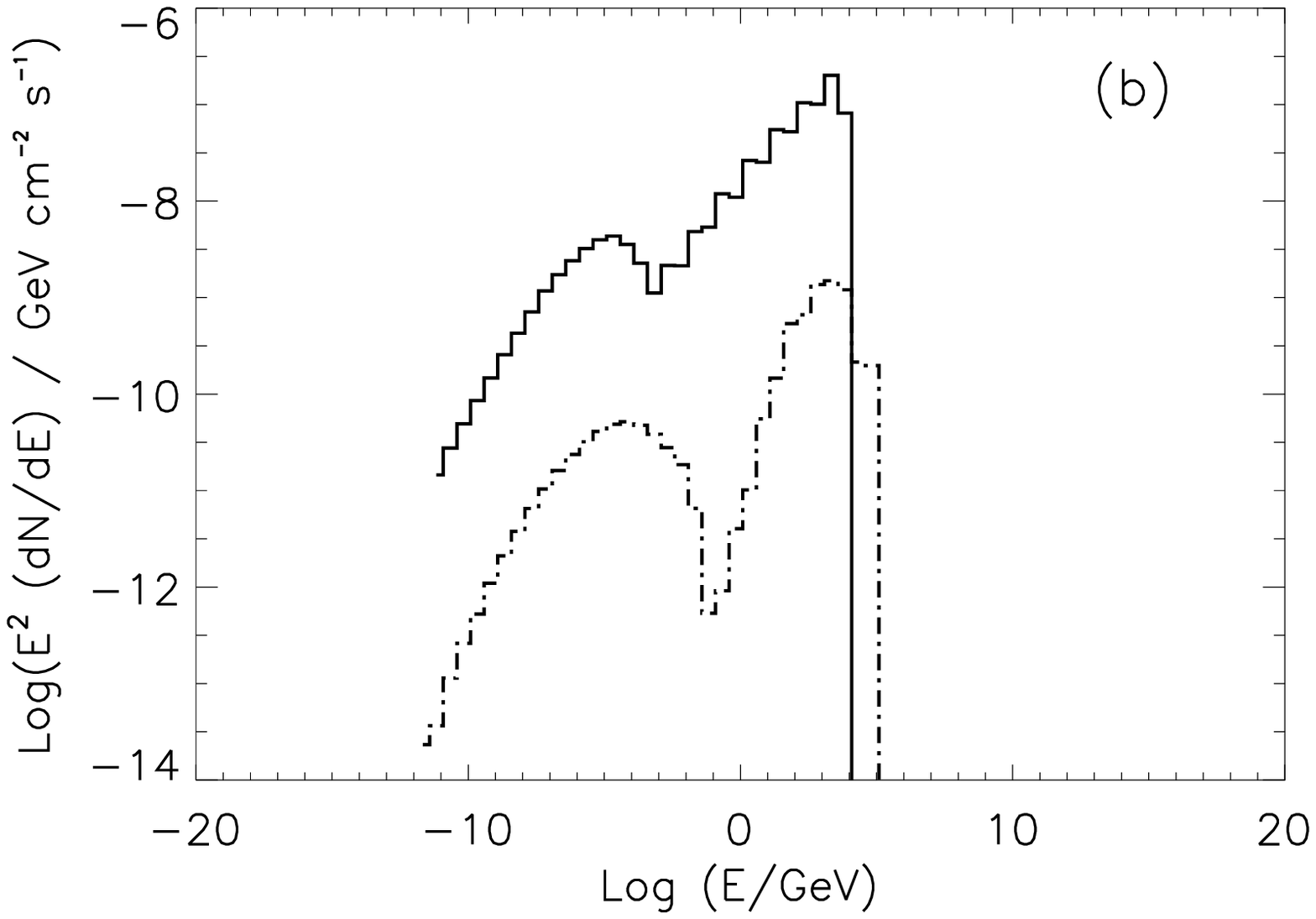,height=3.5in,width=3.7in}}
\vspace{10pt}
\caption{(a) Average emerging cascade spectrum initiated
by Bethe-Heitler pair production. (b) Average emerging
cascade spectrum initiated by $p$- (solid histogram) and
$\mu^\pm$-synchrotron photons (dot-dashed histogram).  Note the
different vertical scale in both figures.}
\label{fig4}
\end{figure}


\section{Application to the April 1997 flare of Mkn~501}

Adding the four components of the cascade spectra in Fig.~4-5 we
obtain the SED shown in Fig.~6 where it is compared with the
multifrequency observations of the 16 April 1997 flare of
Mkn~501.  We use the parametrization of Bednarek \& Protheroe
(1999) to represent the BeppoSAX+OSSE data (thick straight line
at low energies).  The broken power-law simplification
(Eq.~\ref{eq:broken_pl}) used as the target radiation field for
$p\gamma$-interactions and $\gamma\gamma$ interactions is shown
by the chain line.  The 100~MeV upper limit from Catanese et al
(1997) is nearly simultaneous to the TeV-flare observations.  The
TeV-flux, corrected for pair production on the cosmic background
radiation field (Bednarek \& Protheroe 1999) for two different
background models, is shown as the thick curves.

The parameters used for modeling the April 1997 flare are: $D =
12$, $B\approx 20$~G, radius of the emission region
$R_{\rm{blob}}=8 \times 10^{15}$~cm.  For a target photon field
for the $p\gamma$-interactions, and the cascades as given in
Eq.~16, we find a photon energy density of this radiation field
of $u_{\rm{target}} = 60$~GeV/cm$^{-3}$.  The accelerated protons
are assumed to follow a power law $\propto \gamma_p^{-2}$ between
$2\leq \gamma_p \leq \gamma_{p,\rm{max}} = 3\times 10^{10}$, and
in order to fit the emerging cascade spectra to the data a proton
number density of $n_p \approx 250$~cm$^{-3}$, corresponding to an
energy density of accelerated protons of $u_p \approx
11.6$~TeV/cm$^{-3}$ is required.  With a magnetic field energy
density of $u_B \approx 11.7$~TeV/cm$^{-3}$ our model satisfies
$u_{\rm{target}} \leq u_p \approx u_B$ (all parameters are in the
co-moving frame of the jet), confirming that a significant
contribution from inverse-Compton scattering is not expected.

To calculate the total jet luminosity $L_{\rm{jet}}$, measured in the rest
frame of the galaxy,
we adapt the formulae of Bicknell (1994) and Bicknell \& Dopita (1997)
given for the synchrotron self-Compton model to apply for the case of our
proton blazar model. We receive
\begin{eqnarray}
L_{\rm jet} &=& {L^{\rm high}_{\rm obs} \over
D^2 \zeta_p} \left[\chi_p {(\Gamma-1)\over\Gamma} + 1 +
{p_B\over p_p}+ {\zeta_p S^{\rm low}_{\rm obs} \over \zeta_e S^{\rm
high}_{\rm
obs}} \right]
\end{eqnarray}
where $\Gamma = (1 - \beta^2)^{-1} \approx D/2$ is a good
approximation to the Lorentz factor of jets closely aligned to
the line of sight, and
the four terms inside the bracket give the relative
contributions
to the total jet power of cold protons, accelerated protons, magnetic
field,
and accelerated electrons, respectively. The contribution from cold
protons is estimated on the basis of charge neutrality.
$S^{\rm low}_{\rm obs}$ and $S^{\rm high}_{\rm obs}$ are the observed
bolometric fluxes of the low and high energy component, respectively, and
$\zeta_e\approx 1$, $\zeta_p$ are the radiative efficiencies for
electrons and protons.
$L^{\rm high}_{\rm obs} = 4 \pi d^2 S^{\rm high}_{\rm
obs}$ with $d$ the luminosity distance, $p_B=[B^2/(2\mu_0)]/3$ and
\begin{equation}
p_p = {L^{\rm high}_{\rm obs} \over 4 D^2 \zeta_p \Gamma^2 \beta c
\pi R_{blob}^2}
\end{equation}
gives the jet-frame pressure
of relativistic protons that would apply in the absense of energy loss
mechanisms, and
\begin{equation}
\chi_p = {3\over4}\left({m_p \over m_e}{\zeta_p S^{\rm low}_{\rm obs} 
\over \zeta_e S^{
\rm high}_{\rm
obs}}{1 \over{\gamma_e}_1 \ln({\gamma_e}_2/{\gamma_e}_1)} - {1 \over {\gamma_p}_1
\ln({\gamma_p}_2/{\gamma_p}_1)}\right).
\end{equation}
We use this formula to estimate the total jet power to be $\approx 10^{46}$~erg/s,
and we find the contributions
to the total jet power of cold protons, magnetic field,
and accelerated electrons, relative to that of accelerated protons, to
be 0.8, 1, and 0.01, respectively, with a number density of cold
protons $\approx 10^4$~cm$^{-3}$. Thus, for the shock acceleration mechanism
we require more power per particle going into relativistic protons than into electrons.


Fig.~7 shows the dependence of the total jet luminosity on the
Doppler factor $D$ with fixed parameters $t_{\rm{var}} =
12$~hours, $\beta_1=0.5$, and with $B$, $R$, $n_p$ and $u_{\rm
target}$ being of Mkn~501 during flaring. Clearly visible is the
fact that in hadronic models the proton kinetic energy and the
poynting flux dominate the total jet luminosity, while the
electron kinetic energy is only of minor importance. At high Doppler
factors the emission region becomes so large that one needs only
relatively small magnetic fields and proton densities to fit the
observations. In addition, adiabatic losses become small
resulting in a decrease of the required kinetic proton
luminosity.  For example, $B\approx 5$~G and $n_p \approx
10^{-2}$~cm$^{-3}$ are sufficient to fit the Mkn~501 flare for
$D=50$, while for $D=8$ magnetic fields of over 30~G and proton
densities of $n_p \approx 10^{4}$~cm$^{-3}$ are needed. The total jet
luminosity exhibits a minimum of $\approx 10^{46}$erg/s at around
$D\approx 12$, which corresponds to the model we have chosen to
present here.  

Smaller variability time scales in the shock acceleration model
can only be explained by quasi-perpendicular shocks with a large
mean free path (see Fig.~1a), which in turn are only consistent
with the diffusion approximation for low shock speeds. Due to the
size limit of the shock, the maximum energy gain is then limited
to significantly lower values, and one may not reach TeV-photon
energies in the proton synchrotron model, unless the magnetic
field is increased accordingly, which in turn leads to larger jet
luminosities. For example, for $D=10$ and a variability time
scale of 12~hours, at least B=20~G is needed to comply with all
acceleration constraints. For $t_{\rm{var}}=3$~hours, one needs
at least 50~G.  In comparison, leptonic models would give a
minimum jet luminosity of $\approx 10^{44}$~erg/s for this flare
(Ghisellini 1998), about two orders of magnitude lower.  This is
due to the much lower magnetic fields invoked there, and the less
massive particles which drive the kinetic flow.

In the framework of the jet--disk symbiosis (e.g. Falcke \&
Biermann 1995), the jet luminosity should not exceed the total
accretion power $Q_{\rm{accr}}$ for the equilibrium state.
Accretion theory relates the disk luminosity to the accretion
power. Page \& Thorne (1974) give $L_{\rm{disk}} \approx (0.05-0.3)
Q_{\rm{accr}}$.  Disk luminosities for 'typical' radio-loud AGN
lie in the range $L_{\rm{disk}} \approx 10^{44}-10^{48}$~erg/s
with BL~Lac objects tending to the lower end on
average. Specifically, for Mkn~501 there are no emission line
measurements available, and this complicates the evaluation of
the disk luminosity for this object. However, any observed
UV-emission in the flaring stage may put an upper limit on it.
Again only historical data are available here, leaving room for
speculation.  We estimate $L_{\rm{disk}} \approx
10^{43}-10^{44}$~erg/s (Mufson et al 1984, Pian et al 1998), and
obtain for the accretion power, $Q_{\rm{accr}} \approx (3-200)\cdot
10^{43}$~erg/s, at least a factor 5 below the necessary value to
comply with the constraint of the disk--jet symbiosis. Note,
however, that the estimate of $Q_{\rm{accr}}$ is based on
archival non-flaring data from Mkn~501, and we could speculate
that either the disk has pushed more energy into the jet during
TeV-flaring, or that the flaring stage can not be considered as a
steady state.
Also, accretion theory might predict larger
conversion efficiencies of the accretion power into disk radiation
than actually might occur in BL~Lac objects.

It is also instructive to consider the radiative efficiency of
the proposed model in comparison to alternative models. 
Using Eq.~29 we estimate the radiation efficiency of the protons
we find $\zeta_p \approx 0.01$.
This is similar
to the value of $\zeta_p$ in the Proton-Blazar model proposed by
Mannheim (1993), and is fully in agreement with the results
presented by Celotti \& Fabian (1993) on a basis of a sample of
105 sources.
In comparison, leptonic models would give
two orders of magnitude higher radiation efficiencies.

\begin{figure}[h] 
\centerline{\epsfig{file=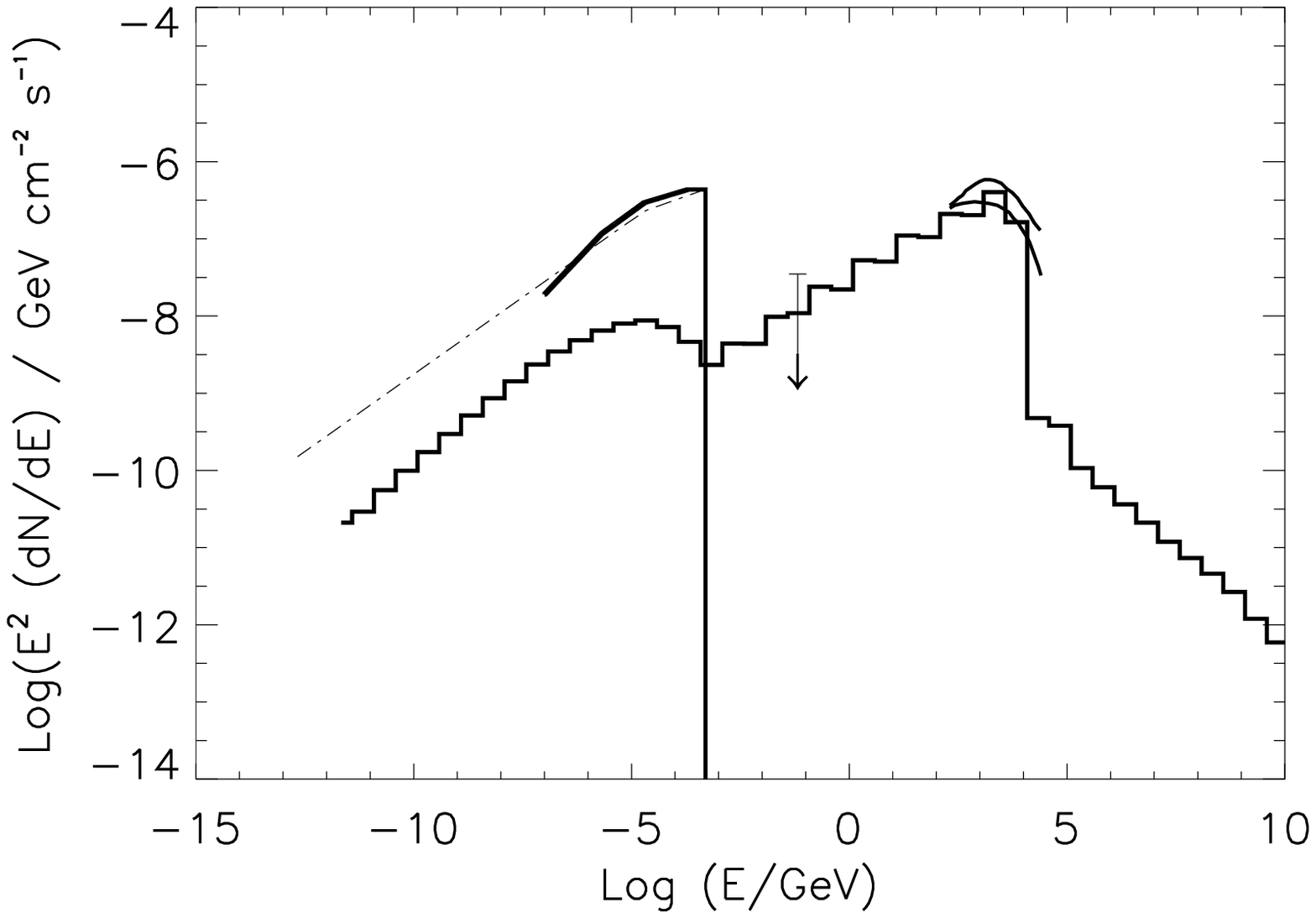,height=4in,width=5in}}
\vspace{10pt}
\caption{ Best-fit model (histogram) in comparison with the data
of the 16 April 1997-flare of Mkn~501. Photon absorption on the
cosmic diffuse background radiation field is not
included. Straight solid lines: parametrization of the observed
synchrotron spectrum (BeppoSAX \& OSSE) and observed TeV-emission
corrected for cosmic background absorption (Bednarek \& Protheroe
1999); the 100~MeV upper limit is from Catanese et al 1997
(observed 9-15 April 1997);
dashed-dotted line: input target spectrum.  }
\label{fig6}
\end{figure}

\begin{figure}[h] 
\centerline{\epsfig{file=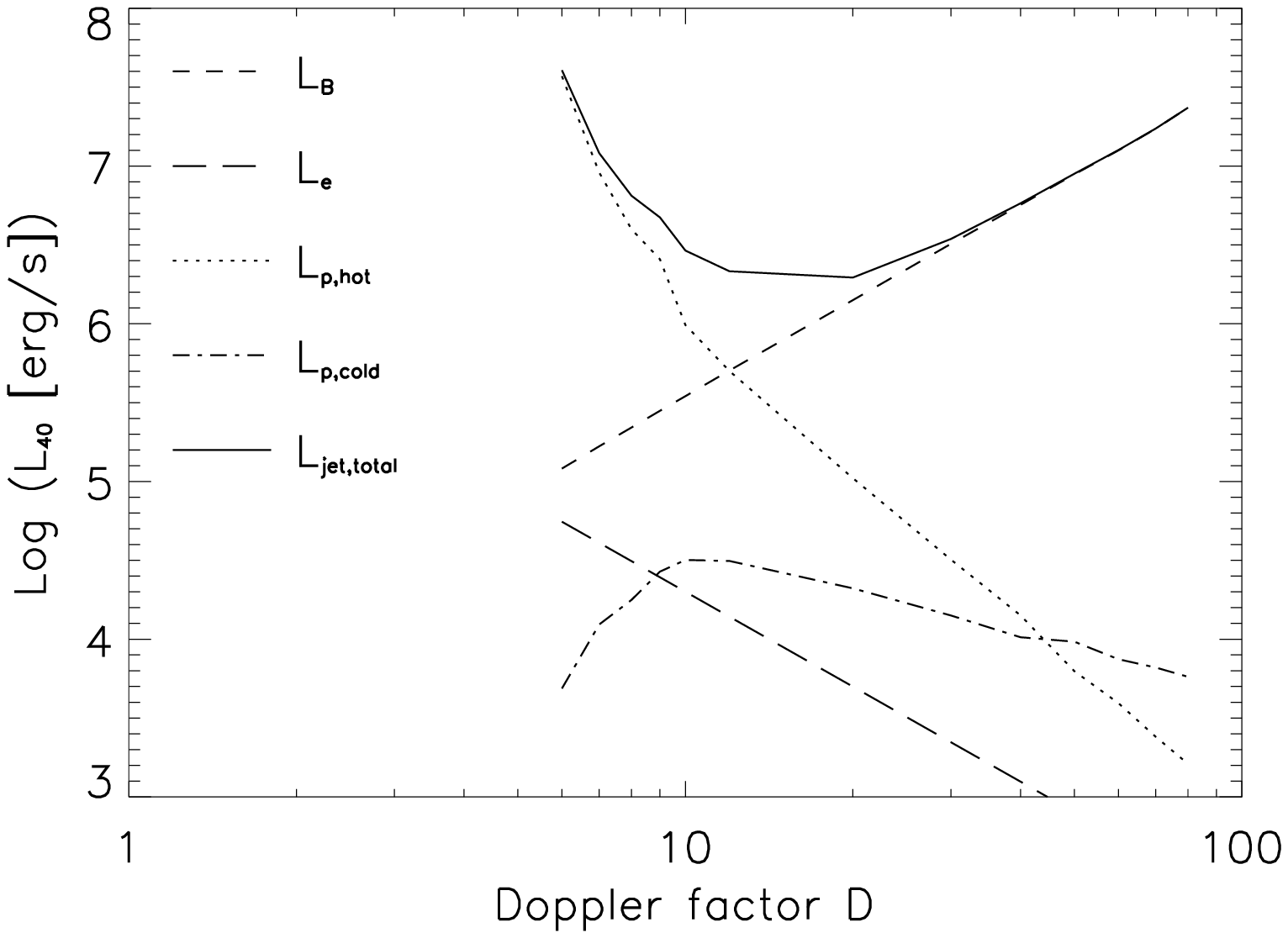,height=3.5in,width=3.7in}}
\vspace{10pt}
\caption{Dependence of total jet luminosity $L_{\rm{jet}} =
L_B+L_{\rm{p,hot}}+L_{\rm{p,cold}}+L_{\rm{e}}$ 
and its different contributions, $L_B$ for the magnetic field,
$L_{\rm{p,hot}}$ and $L_{\rm{p,cold}}$ for the hot and cold
protons and $L_{\rm{e}}$ for the electron contribution, on the 
Doppler factor $D$ for model parameters which reasonably fit the Mkn~501
flare SED. Fixed parameters are
$t_{\rm{var}} = 12$~hours, $\beta_1=0.5$ and $N_p \propto
\gamma_p^{-2}$ for $2\leq \gamma_p \leq \gamma_{p,\rm{max}} = 3
\times 10^{10}$, adjusted parameters are $B$, $R$, $n_p$,
$u_{\rm{target}}$, the synchrotron photon break energy
$\epsilon_b$ and its maximum photon energy
$\epsilon_2$. $L_{jet}$ shows a minimum of $\sim 10^{46}$erg/s
at around $D\approx 12$.  $L_{40} = L/10^{40}$~erg/s.  }
\label{fig4}
\end{figure}

\subsection{Neutrino spectra}

Unlike leptonic models, hadronic blazar models may result in
neutrino emission through the production and decay of charged
mesons, e.g. $\pi^+\to\mu^+ +\nu_{\mu}$ followed by $\mu^+\to e^+ +
\nu_e +\bar{\nu}_{\mu}$. Neutrinos escape without further
interaction, and the predicted neutrino spectrum from Mkn~501
during flaring is shown in Fig.~8.  We calculate the
$\nu$-emission from the source itself, and do not include here
any additional contribution from escaping cosmic rays interacting
while propagating through the cosmic microwave background
radiation (see e.g. Protheroe and Johnson 1996). 
  
The proton injection spectrum is modified by the photo-hadronic
interaction rate which approximately follows a $\gamma_p^{+0.6}$
power-law for proton energies above $\sim 10^{5}$~GeV (see
Fig.~3), where the nucleons interact preferably in the flatter
part of the target photon spectrum ($\epsilon \leq 1.6$~keV) to
produce mesons.  This causes the resulting neutrino spectrum
above $E_\nu \approx 10^5$~GeV to be $dN/dE_{\nu} \propto
E_{\nu}^{-1.4}$, whereas below this energy, the target photon
field for pion production is the steeper part (1.6~keV $\leq
\epsilon \leq 42$~keV), and the corresponding neutrino spectrum
is $dN/dE_{\nu} \propto E_{\nu}^{-1.2}$. At even lower energies a
further flattening is due to pion production by the lowest energy
protons at threshold.  There is a steepening in the spectrum
above $10^9$~GeV which needs some comment here. Neutrinos with
energy $<10^9$~GeV are mostly produced near threshold in the
$\Delta$-resonance region, while the higher energy neutrinos are
mainly produced in the secondary resonant region of the pion
production cross section. $\pi^-$ production is suppressed in the
$\Delta$-resonance, and hence we find a suppression of
$\bar\nu_e$-emission in comparison to $\nu_e$ emission at low
energies, whereas $\pi^+$ and $\pi^-$ multiplicities are
comparable at higher CM energies, leading to roughly equal fluxes
of $\nu_e$ and $\bar\nu_e$.  The effects of $\mu^\pm$-synchrotron
emission show up as a break at $\sim 10^{9}$~GeV in the
observer's frame, whereas $\pi^\pm$-synchrotron emission turns out
to be unimportant in our model.

Another important source of high energy neutrinos is the
production and decay of charged kaons.  In the case of
$p-\gamma$-interactions positively charged kaons are produced,
and are important in the secondary resonance region of the cross
section (M\"ucke et al 1999, M\"ucke et al 2000). They decay in $\sim 64\%$ of all
cases into muons and direct high energy muon-neutrinos. In
contrast to the neutrinos originating from $\pi^\pm$ and
$\mu^\pm$-decay, these muon-neutrinos will not have suffered
energy losses through $\pi^\pm$- and $\mu^\pm$-synchrotron
radiation, and therefore appear as an excess in comparison to the
remaining neutrino flavors at the high energy end ($E_\nu \approx
10^{9}$--$10^{10}$~GeV) of the emerging neutrino spectrum, and in
addition cause the total neutrino spectrum to extend to $\sim
10^{10}$~GeV.

In contrast to previous SPB-jet models in which one expected
equal photon and neutrino energy fluxes (e.g. Mannheim 1993,
1995), our model predicts a peak neutrino energy flux
approximately two orders of magnitude lower than high energy
gamma-rays. This is due to the synchrotron losses of protons
being dominant, leading to gamma-ray emission at the expense of
neutrino emission.  

\begin{figure}[h] 
\centerline{\epsfig{file=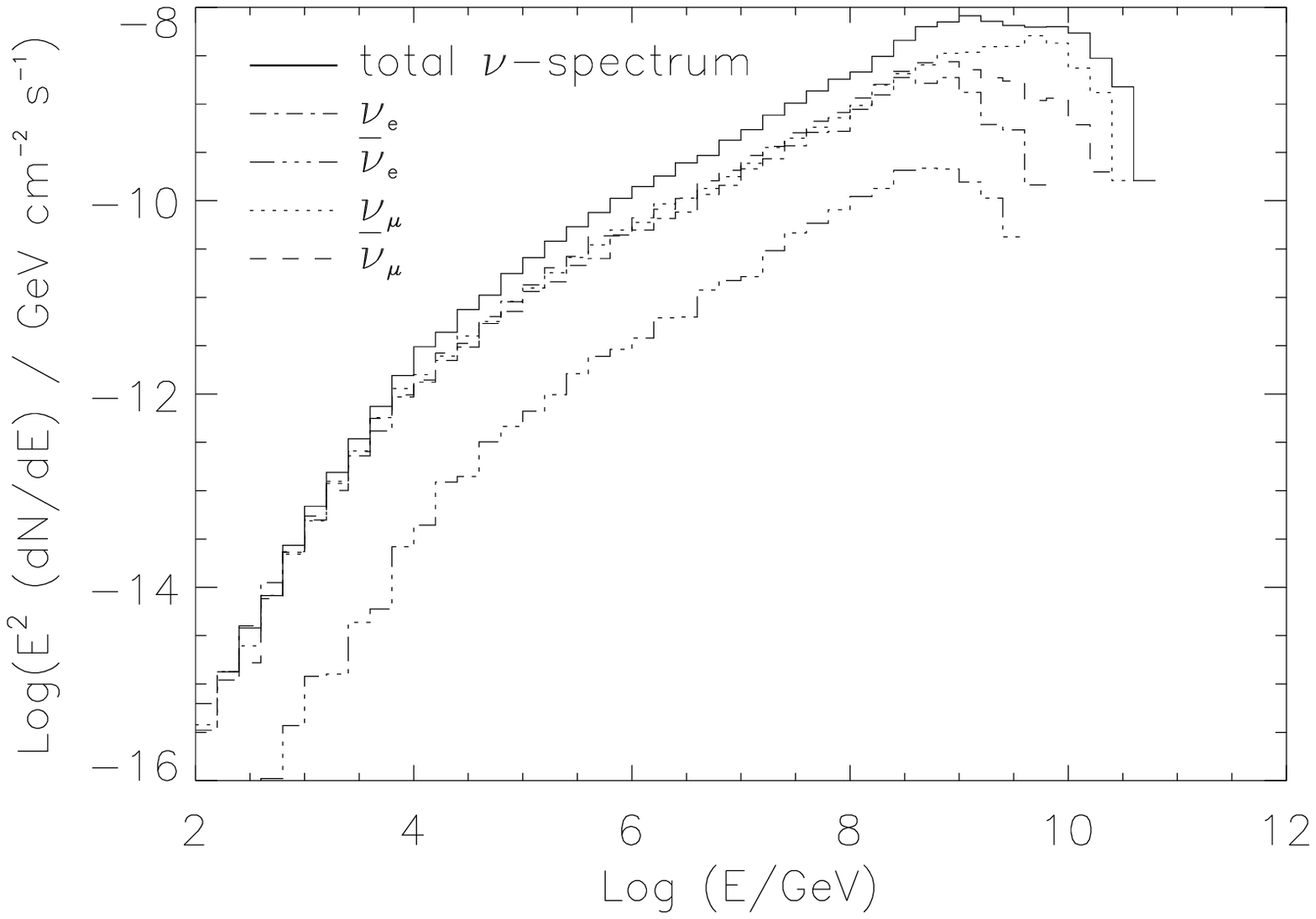,height=4in,width=5in}}
\vspace{10pt}
\caption{Predicted spectra of $\nu_e$, $\bar\nu_e$,
$\nu_\mu$ and $\bar\nu_\mu$ from Mkn~501.  The contribution of
$\nu$s due to pion production of the emerging cosmic rays while
propagating through the cosmic background is not considered
here.}
\label{fig7}
\end{figure}

\section{Conclusions}

This paper describes an application of our newly-developed Monte
Carlo program which simulates a modified version of the
stationary SPB-model.  The Monte Carlo technique allows us to use
exact cross sections, and all important emission processes are
considered here.  As an example, we have used our code to model
the giant April 1997 TeV-flare from Mkn~501. Here, the
TeV-photons are due to synchrotron radiation of the relativistic
protons in the highly magnetized emission region. This proton
synchrotron model was first proposed by M\"ucke \& Protheroe
(1999); a similar model for TeV emission in Mrk~501 has just been
proposed by Aharonian (2000), and his conclusions regarding the
required Doppler factor and magnetic field are very similar to
ours.
Our model departs from the standard SPB model as introduced first
by Mannheim and co-workers mainly in two areas: (i) we use the
{\it{observed}} synchrotron radiation as the target photon
field for $p\gamma$-interactions and pair-synchrotron cascades
assuming it to be produced by co-accelerated electrons, and (ii)
our model takes into account synchrotron radiation from muons and
protons.  The model parameters derived assuming diffusive shock
acceleration of $e^-$ and $p$ in a Kolmogorov turbulence spectrum
are consistent with the X-ray to TeV-data in the flare state.
However, the total jet power we obtain is too large to comply
with the steady-state jet--disk symbiosis scenario, but then we
are not dealing with a stready-state phenomenon.

While the emerging cascade SED initiated by $\pi^0$ decay and
$\pi^\pm$ synchrotron photons turns out to be relatively
featureless, as was also found by, e.g., Mannheim (1993), the
$\mu^\pm$ (see also Rachen 1999, and Rachen \& Mannheim 2000)
and, more importantly, the proton synchrotron radiation and its
cascade produces a double-humped SED as is commonly observed in
flaring blazars. For the present model, we find that proton
synchrotron radiation dominates the TeV emission, while the
contribution of the synchrotron radiation from the pairs,
produced by photon-photon interactions of gamma-rays from the
high energy hump is only minor.  Our model considers the emission
region to be homogeneous.  Inhomogeneities in particle density,
magnetic field, etc. within the source would result in a broader
X-ray and TeV-peak in the SED. This indicates that for Mkn~501 a
homogeneous model of the emission region seems to be appropriate.

Being a hadronic model, our model predicts neutrino emission and
we give the expected neutrino flux of Mrk~501 during
flaring. Comparing our predicted neutrino flux with that for
previous proton blazar models (e.g. Mannheim 1993), we find that
the neutrino output in our model is significantly less than was
previously estimated due to the synchrotron losses dominating the
energy losses of protons, producing synchrotron $\gamma$-rays at the 
expense of $\pi^0$ $\gamma$-rays and neutrinos.

\section*{Acknowledgements}
We thank J\"org Rachen, Alina Donea and Geoff Bicknell for helpful discussions.
This work was supported by a grant from the Australian Research Council.

\end{document}